\documentclass[twocolumn,superscriptaddress,floatfix,prb, aps]{revtex4-1}
\usepackage{graphicx,amsfonts,amssymb,amsmath,hyperref,hypcap,enumerate,enumitem}
\usepackage{xcolor}
\usepackage{soul}
\usepackage{scalerel}
\usepackage{adjustbox,stmaryrd}
\usepackage{cancel}
\usepackage{csquotes}
\usepackage[utf8]{inputenc}
\usepackage{geometry}
\hypersetup{  colorlinks=true, linkcolor=blue, citecolor=red, urlcolor=blue  }

\begin{document}

\title{Interaction and coherence in 2D bilayers}
\author{Jihang Zhu}
\affiliation{Condensed Matter Theory Center and Joint Quantum Institute, Department of Physics, University of Maryland,
College Park, Maryland 20742, USA}
\author{Sankar Das Sarma}
\affiliation{Condensed Matter Theory Center and Joint Quantum Institute, Department of Physics, University of Maryland,
College Park, Maryland 20742, USA}

\begin{abstract}
Two-dimensional electron gas (2DEG) bilayers provide suitable platforms for electronic phases and transitions that are fundamental to both theoretical physics and practical applications in device technology. 
In bilayer systems, the additional pseudospin—representing the layer degree of freedom—enables the emergence of interlayer coherence, which is a direct consequence of the interlayer Coulomb interaction.
This study presents a comprehensive Hartree-Fock (HF) mean-field investigation of the interlayer coherence in 2D bilayers, uncovering ground-state behaviors and temperature-dependent phase transitions that are distinct from single-layer 2DEG. This interlayer coherence signals a spontaneous breaking of the U(1) symmetry in layer pseudospin.
We explore the zero-temperature phase diagrams as a function of the electron density and interlayer separation within the HF formalism.
We also calculate the critical temperature ($T_c$) of the interlayer coherence onset by self-consistently solving the HF gap-like equation. 
We contrast this interlayer coherent phase in electron-electron (e-e) bilayers with the closely related excitonic superfluid phase in electron-hole (e-h) bilayers.  
Though both e-e and e-h bilayers spontaneously break the pseudospin U(1) symmetry, e-h bilayers produce Bardeen-Cooper-Schrieffe-Bose-Einstein condensates (BCS-BEC) crossover intrinsic to the excitons acting as effective bosons or Cooper pairs, whereas the symmetry-broken phase in e-e bilayers is akin to the XY or easy-plane pseudospin ferromagnetism.
Using the same system parameters and a similar theoretical framework, we find that $T_c$ of the interlayer coherent phase in e-e bilayers is about one-third of that in exciton condensates, suggesting a weaker interlayer coherence in e-e bilayers.
In addition, we examine the effect of a weak interlayer tunneling on the interlayer coherence order parameter, drawing parallels with the influence of an effective in-plane magnetic field on the XY pseudospin ferromagnetism.
Our findings provide a comparative theoretical framework that bridges the gap between the interlayer coherence physics in e-e and e-h bilayers, contributing to a unified understanding of phase transitions in low-dimensional electron/hole systems and establishing in particular the same universality class for interlayer phase coherence in both e-e and e-h bilayers.
\end{abstract}

{\let\newpage\relax\maketitle}

\setcounter{page}{1} 

\section{Introduction}
During the last 50 years, the fabrication and exploration of two-dimensional electron gas (2DEG) systems has been crucial in expanding our understanding of quantum mechanical phenomena in condensed matter physics.\cite{Ando_2D_1982, sarma2008perspectives} These systems, fundamental to the functionality of contemporary electronic devices such as Si MOSFETs and GaAs HEMTs, exhibit rich many-body quantum phases under various conditions, for example at low electron densities or in a strong magnetic field where Coulomb interactions dominate.
The most famous such interacting phase is the fractional quantum Hall effect,\cite{sarma2008perspectives} but many other correlation-induced phases, both in finite and zero magnetic fields, have been predicted and sometimes observed, such as the Wigner crystal,\cite{Wigner_1934} the Bloch ferromagnet,\cite{Bloch_1929} spin/valley/charge density waves, etc.

The Bardeen-Cooper-Schrieffer (BCS) theory \cite{BCS_micro, BCS} describes the pairing of two electrons into a Cooper pair as a result of effective attraction mediated by phonons. An analogous pairing mechanism, induced directly by the Coulomb interaction, arises between an electron and a hole, giving rise to the excitonic state in semimetals or semiconductors. \cite{Keldysh_exciton_1965, Kohn_EI_1967, Keldysh_exciton_1968, kozlov_exciton_1965, Halperin_excitonic_1968}
Even though the original concept of an exciton\cite{Frenkel_exciton_1931, Wannier_exciton_1937, Knox_exciton_1963} describes the bound state of an electron and a hole, which are generated optically in the conduction and valence bands of homogeneous semiconductors, this kind of exciton is an excited state.
Excitons formed in e-h bilayers are, however, the ground state of the bilayer system.  We only consider such e-h bilayer excitons and the corresponding excitonic ground states in the current work. 
The nature of these excitonic states is determined by the pairing strength. 
In the weak pairing limit, excitons are loosely bound with radii larger than the average distance between electrons and holes. In this regime, the original semimetallic or semiconducting phase is unstable for an arbitrarily weak e-h attraction, akin to how the normal Fermi surface of a metal is susceptible to the Cooper-pair formation under any weak e-e attraction in the BCS theory.
In the strong pairing limit, where excitons are tightly bound with smaller radii, excitons behave like a weakly repulsive dilute Bose gas, forming the Bose-Einstein condensates (BEC) at low temperatures.
In either limit, as long as  the electrons and holes remain spatially separated, excitons condense into a ground state which is a BCS or BEC superfluid. Such a superfluid is not a regular superconductor since excitons are electrically neutral, thus this ground state is sometimes called the \enquote*{excitonic insulator}; It also differs from a regular insulator, since this excitonic ground state constitutes a neutral superfluid.
We use the terminology \enquote*{exciton condensates}  for the $T=0$ strongly-bound BEC or weakly-bound BCS superfluid of 2D e-h bilayer excitons.

Advances in quantum well bilayers, such as InAs-GaSb \cite{RuiRuiDu_exciton_2017} and GaAs–AlGaAs\cite{Croxall_CoulombDrag_2008, DasGupta_GaAsQW_2008, Seamons_QW_2007} heterostructures, have enabled the spatial separation of electrons and holes, and therefore the creation of more stable and controllable e-h bilayers.
These developments have facilitated the observations of stable excitonic states in equilibrium. \cite{Lozovik_exciton_1975, Shevchenko_exciton_1976, Naveh_exciton_1996, Datta_exciton_1985, XZhu_exciton_1995, XZhu_exciton_1996, Butov_exciton_2003, RuiRuiDu_exciton_2017,Nguyen_ehbilayer_2023}
In these quantum well bilayers, particularly, exciton condensation has been predicted and observed in a strong magnetic field perpendicular to the 2D layers.\cite{sarma2008perspectives_Girvin, sarma2008perspectives_Eisenstein, Eisenstein_BEC_2004, Eisenstein_exciton_2014, Eisenstein_2DEGinB_2004, Tutuc_GaAs_exciton_2004, Eisenstein_science_2004}
The most stable exciton condensates manifest at a total Landau level filling factor of $\nu=1$, where each layer's lowest Landau level is half-filled.
The emergence of these exciton condensates has been further reinforced by recent breakthroughs in 2D material fabrications,\cite{JIALi_excitonGraphene_2017, XLiu_excitonGraphene_2017} including double-layer graphene,\cite{JIALi_supersolid_2023,CDean_exciton_2022} double bilayer graphene and double-layer transition metal dichalcogenides.
Graphene double layers,\cite{CDean_exciton_2022} in particular, have demonstrated the potential for a BCS-BEC crossover\cite{Meera_BCS-BEC, Randeria_BCS-BEC, Chen_BCSBEC_2005, Bourdel_BCSBEC_ColdAtom_2004, Regal_BCSBEC_ColdAtom_2004, Bartenstein_BCSBEC_ColdAtom_2004, Zwierlein_BCSBEC_ColdAtom_2004, Ries_BCSBEC_ColdAtom_2015, Puneet_BCSBEC_ColdAtom_2018, Shahar_BCSBEC_2017, Nakagawa_BCSBEC_2021, RuiRuiDu_exciton_2017, LZhu_QICoulombDrag_2023}—a continuum between BCS-type weak coupling and BEC-type strong coupling regimes—in condensed matter systems.

In contrast to quantum Hall bilayers, exciton condensates in e-h bilayers without an external magnetic field—hereafter simply referred to as e-h bilayers—have remained elusive \cite{RuiRuiDu_exciton_2017} until the recent observation of enhanced interlayer tunneling anomaly in double bilayer graphene heterostructures, \cite{Tutuc_ehbilayer_2018} signaling the first indication of such condensates.
Considerable obstacles in experimental realization of e-h bilayer exciton condensates have been the difficulties in eliminating impurities and in approaching the BEC limit. The latter involves electrostatically doping carriers to a lower density, minimizing interlayer tunneling and simultaneously reducing the distance between layers.
Regardless of the specific pairing mechanism,
these bilayer systems, with or without an external magnetic field, doped with electrons or holes, share a common core: the spontaneous interlayer coherence associated with an excitonic superfluid arising from Coulomb interactions.

The relative stability of exciton BEC in quantum Hall bilayers at $\nu=1$ can be attributed to the unique characteristics of Landau levels. Electrons in Landau levels bypass the cost of kinetic energies, thus enhancing interaction-dominated symmetry-broken phases.
Furthermore, the quantum Hall bilayer is distinguished by the exact particle-hole symmetry inherent in the Landau level around the half-filling. 
At $\nu=1$, despite both layers being electron-doped and half-filling the lowest Landau level, the exact particle-hole symmetry transforms electrons into holes, and vice versa—quantum Hall bilayers at $\nu=1$ can be equivalently considered as e-e, h-h or e-h bilayers, and thus equivalently describable as an XY pseudospin ferromagnet in the e-e or h-h picture\cite{AHMacDonald_QHF_2000, Fertig_QHBilayer_1989, KunYang_QHbilayer_1996, Boebinger_DQW_1990, Yoshioka_ehfluid_1990, Ezawa_eeQH_1993, KunYang_QHbilayer_1994, Moon_QHbilayer_1995, Kyriienko_2015} or as a BEC superfluid in the e-h picture.\cite{Eisenstein_BEC_2004, Eisenstein_science_2004}
However, the absence of exact particle-hole symmetry, due to differing energy dispersions of electrons and holes, in e-h bilayers prevents a direct mapping between BEC and pseudospin ferromagnetism. 
This naturally raises an essential question: 
How do the nature and characteristics of interlayer coherence differ between e-h and e-e bilayers at zero magnetic field?  Are they the same or different in the absence of particle-hole symmetry?
In this paper, we address this question in depth using the Hartree-Fock (HF) mean-field theory, which has also been used extensively to study the quantum Hall bilayer exciton condensation phenomena. 

Though the comparison between BCS superconductivity and exciton condensates has a long history, there has been no in-depth discussion comparing the pairings in e-h and e-e bilayers.
Closely related, e-e and e-h bilayers both host spontaneous interlayer coherence as a result of repulsive Coulomb interactions between electrons (or attraction between electrons and holes).
In spite of the same Coulombic nature, the interlayer coherence in these two systems exhibit distinct properties. In e-h bilayers, the formation of exciton condensates features a gap opening, while the interlayer coherence in e-e bilayers is more akin to an XY pseudospin ferromagnetism.\cite{XYbilayer_Girvin_2000}
As previously discussed in depth in Refs [\onlinecite{LZheng_doubleQW_1997},\onlinecite{DasSarma_doubleQW_1998}], the layer index introduces a pseudospin, of which the $z$-component maps to the classical layer index.
The pseudospin-ordered state underscores the quantum mechanical subtleties of pseudospin-orientation-dependent energies: the XY easy-plane pseudospin ferromagnetic state is energetically favorable
due to the absence of Hartree contribution.
A profound distinction arises between the symmetry properties of spin and pseudospin. Whereas conventional spin system exhibits SU(2) symmetry, ensuring energy invariance under any rotation of spin orientation, the symmetry of the pseudospin-ordered state in bilayer systems is reduced to U(1),\cite{XGWen_QHB_1992, Nakajima_QHF_1997} which only guarantees energy invariance under rotations in the plane perpendicular to the effective magnetic field. This reduction in symmetry from SU(2) to U(1) reflects the underlying order imposed by the difference between the intralayer and interlayer Coulomb interactions, which is unique to the bilayer systems. 
This pseudospin U(1) symmetry captures the essence of interlayer coherence, akin to the phase coherence in BCS superconductors. 

In this work we focus on the zero-field e-e bilayers, which have been much less studied than the corresponding well-studied e-h bilayers although both manifest similar U(1) symmetry breaking associated with spontaneous interlayer coherence, which physically implies the system developing interaction-induced interlayer tunneling in spite of the absence of any single-particle interlayer tunneling.
Specifically, we consider a bilayer structure, composed of two 2DEG layers confined in the $xy$-plane and separated by a distance $d$ in the $z$-direction. 
Such a bilayer structure introduces new physics:\cite{SDS_WignerSupersolid_2006} in addition to 2D carrier density in each layer, controlling the kinetic energy and the intralayer interaction strength, the layer separation provides a new length and interaction scale for interlayer correlations. 
We study the interlayer coherence in such an e-e bilayer using the HF theory, and for completeness, we compare its finite-temperature properties with e-h bilayers using the same system parameters.
We pedagogically start in Sec.~\ref{Sec_GSAnsatze} by introducing four distinct ground-state ansatz:
the spin and pseudospin unpolarized phase $S_0$, the spin polarized but pseudospin unpolarized phase $S_1$, the spin polarized and interlayer coherent phase with pseudospin in the $xy$-plane $S_2$ and the spin polarized and interlayer coherent phase with pseudospin in $z$-direction $S_3$.
We focus exclusively on the spin polarized case in interlayer coherent phases, primarily because spin plays no role in the XY pseudospin ferromagnetic state if the density is lower than the critical value of the ferromagnetic instability ($r_s > 2$) and the spin-orbit coupling is absent—the system is effectively spinless. Therefore, our findings regarding the interlayer coherence remain equally applicable to spinless itinerant electrons, with the only difference being the absence of spin polarization in each layer. 
The fact that a 2DEG has an exchange-driven spin polarization transition at low densty (for $r_s \sim 2$)\cite{Bloch_1929} is well-known, and our spin polarization transition results are for the sake of completeness only.

In Sec.~\ref{Sec_zeroT}, we study the interplay of kinetic, Hartree and exchange energies, by mapping out the zero-temperature phase diagrams as a function of electron density and interlayer separation.
We partition the discussion into two subsections to consider scenarios of equal layer densities (Sec.~\ref{subsec_phase_equaln}) and unequal layer densities (Sec.~\ref{subsec_unequaln}), both conditions being experimentally adjustable through the application of dual gate voltages.
Interestingly, within the $S_1$ and $S_2$ regimes in the HF energy landscape, our findings suggest a ground state that favors a pseudospin partially polarized phase: both pseudospin-up and pseudospin-down are, unequally, occupied. This contrasts with a fully polarized phase where pseudospin alignment would be exclusively in the $x$-direction (actually any direction in the $xy$-plane because of the U(1) symmetry).

In Sec.~\ref{sec_finiteT}, we extend our study to finite temperatures, focusing on bilayers with equal layer densities. We provide an in-depth investigation of the critical temperature $T_c$ for the interlayer coherent phase ($S_2$) in Sec.~\ref{subsec_TcS2}, drawing comparisons to the exciton condensation in e-h bilayers with the same system parameters in Sec.~\ref{subsec_Tcexciton}. 
The trends of $T_c$, as a function of density and interlayer separation, of e-e bilayers qualitatively deviate from that of e-h bilayers. The magnitude of $T_c$, on the other hand, in e-e bilayers is approximately one-third of that in e-h bilayers, suggesting a weaker interlayer coherence in e-e bilayers.
Both the XY ferromagnetism and the exciton condensates undergo, in principle, a finite-temperature Berezinskii–Kosterlitz–Thouless (BKT) transition, and belong to the same universality class of phase transitions.

Section~\ref{sec_tunneling} investigates the influence of a weak interlayer tunneling on the interlayer coherence order parameter. 
The presence of any finite interlayer single-particle tunneling explicitly breaks the U(1) symmetry and pins the easy-plane pseudospin ferromagnetism along the $x$-direction.
This analysis provides insights into the phase coherence and stability under perturbations that mimic the influence of an external magnetic field to physical spins.

Finally, we conclude with a discussion in Sec.~\ref{sec_discussion}. We comment on possible experimental signatures of the interlayer coherence in e-e bilayers and limitations of the HF mean-field theory.

In addition, we provide in Appendix~\ref{Appendix_integrals} some useful summations and integrals that are used in our theory, in Appendix~\ref{HF_energy_T0} the HF energy plots of the four competing ground states $S_0, S_1, S_2, S_3$, and in Appendix~\ref{sec_appendixC} the $T_c$ of the spin polarized phases in single 2DEG and three-dimensional electron gas (3DEG) systems for completeness.

\section{Ground State Ansatz}
\label{Sec_GSAnsatze}
\begin{figure}
\centering
\includegraphics[width=1.0\columnwidth]{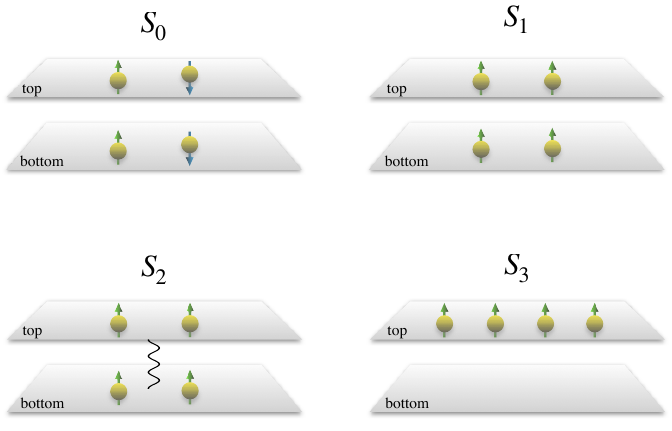}
\caption{\label{fig_GSs} {
Schematically show the four competing ground states we consider: the spin and pseudospin unpolarized phase $S_0$, the spin polarized but pseudospin unpolarized phase $S_1$, the spin polarized interlayer coherent (pseudospin polarized) phase with the pseudospin in the $xy$-plane $S_2$ (the XY order) and with the pseudospin in the $z$-direction $S_3$ (the Ising order).
In the $S_3$ phase, all electrons spontaneously move to one layer creating a charge order.
The interlayer coherence is indicated by the wiggle.
(The $S_\xi$ phase mentioned in the main text consists of both $S_2$ and $S_3$ phases.)
  }}
\end{figure}

We start from the characterization of four distinct ground states that emerge in the study of interlayer coherence in e-e (or h-h) bilayers. We designate these states with concise labels for ease of reference throughout our discussion:
\begin{itemize}[leftmargin=*]
\item $S_0$ phase: The spin and pseudospin unpolarized state.
\item $S_1$ phase: The spin polarized but pseudospin unpolarized state.
\item $S_\xi$ phase: The spin polarized interlayer coherent (pseudospin polarized) states. The pseudospin polarization in these states can be oriented differently.
    \begin{itemize}[leftmargin=*]
    \item $S_2$ phase: The pseudospin is oriented within the $xy$-plane (the XY order).
    \item $S_3$ phase: The pseudospin is oriented in the $z$-direction (the Ising order).
    \end{itemize}
\end{itemize}
$S_2$ and $S_3$ phases both break the U(1) layer symmetry and
correspond to SP-SY and SP-MO phases in Ref.~[\onlinecite{LZheng_doubleQW_1997},\onlinecite{DasSarma_doubleQW_1998}], respectively.
Figure~\ref{fig_GSs} schematically illustrates these four competing ground states.

\section{Zero temperature Phase diagrams}
\label{Sec_zeroT}
In this section, we provide zero-temperature phase diagrams of e-e bilayers.
We first show the HF Hamiltonian and the formula of HF total energy, followed by explicit HF energy expressions of $S_0$, $S_1$, $S_2$ and $S_3$ phases in Sec.~\ref{subsec_HFenergy}. 
We then present phase diagrams with respect to electron density (or dimensionless inter-electron distance $r_s$) and interlayer separation $d$ for equal layer densities\cite{LZheng_doubleQW_1997,DasSarma_doubleQW_1998} in Sec.~\ref{subsec_phase_equaln} and for unequal layer densities in Sec.~\ref{subsec_unequaln}.


The Hamiltonian of a 2D bilayer consists of the kinetic part $\hat{\mathcal{H}}_0$ and the Coulomb interacting part $\hat{\mathcal{V}}$,
\begin{equation}
\hat{\mathcal{H}} = \hat{\mathcal{H}}_0 + \hat{\mathcal{V}}.
\end{equation}
Represented in the second quantization form,
\begin{equation}
\begin{split}
\hat{\mathcal{H}}_0 = \sum\limits_{\mathbf{k},l, \sigma} &\varepsilon_{0,l}(\mathbf{k}) c^\dagger_{l \sigma \mathbf{k}} c_{l \sigma \mathbf{k}}, \\
\end{split}
\end{equation}
where $l=t,b$ label top and bottom layers and $\sigma = \uparrow, \downarrow$ label spins.
For free electron gases, $\varepsilon_{0,l}(\mathbf{k}) = \hbar^2 k^2/2m_l^*$, and $m_l^*$ is the effective mass. 
For simplicity, we assume $m^*_t = m^*_b = m^*$ throughout the rest of the paper. It is straightforward to generalize the case to unequal effective masses $m^*_t \neq m^*_b$, which will only quantitatively change the results we present in this paper.
The interacting part is
\begin{equation}
\begin{split}
\hat{\mathcal{V}} = \frac{1}{2A} 
\sum\limits_{\substack{\mathbf{k},\mathbf{k}',\mathbf{q} \\
l,l',\sigma,\sigma'}} V_{\mathbf{q}}^{ll'} c^\dagger_{l \sigma, \mathbf{k}+\mathbf{q}} c^\dagger_{l' \sigma', \mathbf{k}'-\mathbf{q}} c_{l' \sigma',\mathbf{k}'} c_{l \sigma,\mathbf{k}},
\end{split}
\end{equation}
where $A = A_t = A_b$ is the system area.
The 2D Coulomb potentials are different for the electrons in the same layer and in different layers,
\begin{equation}
V^{ll'}_{\mathbf{q}} = V^S_{\mathbf{q}} \delta_{ll'} + V^D_{\mathbf{q}} (1-\delta_{ll'}),
\end{equation}
where
\begin{equation}
V^S_{\mathbf{q}} = \frac{2\pi e^2}{\epsilon q}, \quad
V^D_{\mathbf{q}} = \frac{2\pi e^2}{\epsilon q} e^{-qd},
\end{equation}
$d$ is the layer separation and $\epsilon$ is the averaged dielectric constant of the surrounding medium.
We note that the interaction is spin-independent (i.e., SU(2) symmetric), but layer-dependent (i.e., U(1) symmetric in pseudospin layer index).

The HF mean-field form of the interacting part $\hat{\mathcal{V}}$ is
\begin{equation}
\begin{split}
\hat{\mathcal{V}}_{\rm HF} &= \sum\limits_{\mathbf{k},l,\sigma} \Big[ \left( V_{H,l} + V_{x,l}^{\sigma \sigma}(\mathbf{k})
\right) c^\dagger_{l \sigma \mathbf{k}} c_{l \sigma \mathbf{k}} \\
& + V_{x,l}^{\sigma \bar{\sigma}}(\mathbf{k}) c^\dagger_{l \sigma \mathbf{k}} c_{l \bar{\sigma} \mathbf{k}} \Big]
-\sum\limits_{\mathbf{k},\sigma,\sigma'}\Delta^{\sigma \sigma'}_{\mathbf{k}} c^\dagger_{t \sigma \mathbf{k}} c_{b \sigma' \mathbf{k}} 
- h.c..
\end{split}
\end{equation}
We have explicitly separated $\hat{\mathcal{V}}_{\rm HF}$ into Hartree potentials $V_{H,l}$, intralayer exchange potentials between the same spin $V_{x,l}^{\sigma \sigma}(\mathbf{k})$, intralayer exchange potentials between opposite spins $V_{x,l}^{\sigma \bar{\sigma}}(\mathbf{k})$ and interlayer exchange potentials $\Delta_{\mathbf{k}}^{\sigma \sigma'}$.

In the matrix form with spinor basis $(c_{t \uparrow \mathbf{k}}$  $c_{t \downarrow \mathbf{k}}$ $c_{b \uparrow \mathbf{k}}$ $c_{b \downarrow \mathbf{k}})^T$, the HF Hamiltonian $\hat{H}_{\rm HF}(\mathbf{k}) = \hat{H}_0(\mathbf{k}) + \hat{V}_{\rm HF}(\mathbf{k})$ is given by
\begin{widetext}
\begin{equation}
\label{Eq_HMF}
\hat{H}_{\rm HF}(\mathbf{k}) = 
\begin{pmatrix}
\varepsilon_{0}(\mathbf{k})+V_{H,t}+V^{\uparrow \uparrow}_{x,t}(\mathbf{k}) & V^{\uparrow \downarrow}_{x,t}(\mathbf{k}) & -\Delta_\mathbf{k}^{\uparrow \uparrow} & -\Delta_\mathbf{k}^{\uparrow \downarrow} \\
V^{\downarrow \uparrow}_{x,t}(\mathbf{k}) & \varepsilon_{0}(\mathbf{k})+V_{H,t}+V^{\downarrow \downarrow}_{x,t}(\mathbf{k}) & -\Delta_\mathbf{k}^{\downarrow \uparrow} & -\Delta_\mathbf{k}^{\downarrow \downarrow} \\
-\Delta_\mathbf{k}^{*\uparrow \uparrow} & -\Delta_\mathbf{k}^{*\downarrow \uparrow} & \varepsilon_{0}(\mathbf{k})+V_{H,b}+V^{\uparrow \uparrow}_{x,b}(\mathbf{k}) & V^{\uparrow \downarrow}_{x,b}(\mathbf{k}) \\
-\Delta_\mathbf{k}^{*\uparrow \downarrow} & -\Delta_\mathbf{k}^{*\downarrow \downarrow} & V^{\downarrow \uparrow}_{x,b}(\mathbf{k}) & \varepsilon_{0}(\mathbf{k})+V_{H,b}+V^{\downarrow \downarrow}_{x,b}(\mathbf{k})
\end{pmatrix}.
\end{equation}
\end{widetext}
The electrostatic Hartree potentials are
\begin{equation}
\label{Eq_VH}
\begin{split}
V_{H,l} &= \frac{2\pi e^2 d}{\epsilon} n_l,
\end{split}
\end{equation}
where $n_l$ is the carrier density in layer $l$,
\begin{equation}
\begin{split}
n_l = \frac{1}{A}\sum\limits_{\mathbf{k},\sigma} \rho^{\sigma \sigma}_{ll}(\mathbf{k}),
\end{split}
\end{equation}
and $\rho(\mathbf{k})$ is the density matrix with matrix elements
\begin{equation}
\rho^{\sigma \sigma'}_{ll'}(\mathbf{k}) = \langle c^\dagger_{l'\sigma' \mathbf{k}} c_{l\sigma \mathbf{k}} \rangle.
\end{equation}
The expectation is taken in the ground state.
The intralayer exchange potentials are
\begin{equation}
V^{\sigma \sigma'}_{x,l}(\mathbf{k}) = -\frac{1}{A} \sum\limits_{\mathbf{k}'} V^{S}_{\mathbf{k}-\mathbf{k}'} 
\rho^{\sigma \sigma'}_{ll}(\mathbf{k}'),
\end{equation}
and the interlayer exchange terms are
\begin{equation}
\Delta_{\mathbf{k}}^{\sigma \sigma'} = \frac{1}{A} \sum\limits_{\mathbf{k}'} V^{D}_{\mathbf{k}-\mathbf{k}'} \rho^{\sigma \sigma'}_{tb}(\mathbf{k}').
\end{equation}
The HF total energy is the sum of all contributions,
\begin{equation}
\label{Eq_Etot}
\begin{split}
E_{\rm tot} = &\sum\limits_{\mathbf{k}, l, \sigma} \big( \varepsilon_{0}(\mathbf{k}) + \frac{1}{2} V_{H,l}\big)\rho_{ll}^{\sigma \sigma}(\mathbf{k}) \\
&+ \frac{1}{2} \sum\limits_{\mathbf{k}, l, \sigma, \sigma'} V^{\sigma \sigma'}_{x,l}(\mathbf{k}) \rho_{ll}^{\sigma' \sigma}(\mathbf{k}) \\
&- \frac{1}{2} \sum\limits_{\mathbf{k}, \sigma, \sigma'}  \big[ \Delta_\mathbf{k}^{\sigma \sigma'} \rho_{bt}^{\sigma' \sigma}(\mathbf{k}) + c.c. \big].
\end{split}
\end{equation}
The first (noninteracting single-particle) term includes the kinetic energy and electrostatic Hartree energy, the second and the last (interacting) terms are intralayer and interlayer exchange energies, respectively, arising from the Coulomb coupling.

In subsequent analyses, we use the effective Bohr radius $a^*$ and the effective Rydberg Ry$^*$, defined as
\begin{equation}
a^* = \frac{\epsilon \hbar^2}{m^* e^2}, \quad {\rm Ry}^* = \frac{e^2}{2a^* \epsilon} = \frac{\hbar^2}{2m^* (a^*)^2},
\end{equation}
as fundamental units of length and energy, respectively.
In the semiconductor double quantum well structure GaAs-AlGaAs, $\epsilon = 12.5$, $m^*=0.07m_e$,\cite{shklovskii2013} $a^* = 98.3 \mathring{A}$ and 
Ry$^* \approx 5.5$ meV. 
In a 2DEG, the average distance between electrons is quantified by the dimensionless length scale $r_s$, which is related to the density $n$ by
\begin{equation}
r_s a^* = \frac{1}{\sqrt{\pi n}}.
\end{equation}
Our calculations are presented in terms of both dimensionless quantities
$(r_s, \tilde{d})$, where $\tilde{d} = d/a^*$, and experimentally measurable parameters $(n, d)$ using GaAs quantum well conduction band parameters.
Some minor modifications are necessary for graphene, where the kinetic energy term is linearly dispersing, but our results remain qualitatively valid.

\subsection{The HF energy}
\label{subsec_HFenergy}
Based on Eqs.~(\ref{Eq_VH}-\ref{Eq_Etot}), we explicitly show HF energies of the four competing ground states $S_0$, $S_1$, $S_2$ and $S_3$ in this subsection.
An interlayer coherent state emerges from a superposition of electron states residing in the top and bottom layers. The eigenstates of this pseudospin polarization can be represented as a linear combination:
\begin{equation}
\begin{split}
|\xi \rangle &= \alpha |t \rangle + \beta |b \rangle, \\
|\bar{\xi} \rangle &= \beta |t \rangle - \alpha |b \rangle,
\end{split}
\end{equation}
where $\alpha$ and $\beta$ can be taken as real numbers without loss of generality, and $\alpha^2+\beta^2=1$ satisfying the normalization condition. 
The symmetric state $|\xi\rangle$ and anti-symmetric state $|\bar{\xi} \rangle$ correspond to the lower and higher eigenenergies, respectively.

\subsubsection{HF energy of the spin and pseudospin unpolarized state $S_0$}
The wavefunction of the spin and pseudospin unpolarized state $S_0$ can be written as
\begin{equation}
\begin{split}
|S_0 \rangle &= \prod\limits_{k \leq k_F} c^\dagger_{\xi \uparrow \mathbf{k}} c^\dagger_{\xi \downarrow \mathbf{k}} c^\dagger_{\bar{\xi} \uparrow \mathbf{k}} c^\dagger_{\bar{\xi} \downarrow \mathbf{k}} |0 \rangle \\
&= \prod\limits_{k \leq k_F} c^\dagger_{t \uparrow \mathbf{k}} c^\dagger_{t \downarrow \mathbf{k}} c^\dagger_{b \uparrow \mathbf{k}} c^\dagger_{b \downarrow \mathbf{k}} |0 \rangle,
\end{split}
\end{equation}
where $|0\rangle$ is the vacuum state. There are four equal Fermi surfaces with Fermi momentum $k_{F} = \sqrt{\pi n}$.
The densities and HF potentials of $S_0$ state are
\begin{equation}
\begin{split}
 &n_t = n_b = \frac{n}{2}, \\
 &\rho_{ll'}^{\sigma \sigma'}(\mathbf{k}) = \delta_{ll'} \delta_{\sigma \sigma'} f_\mathbf{k}, \\
 &V_{H,t} = V_{H,b} = \frac{\pi e^2 dn}{\epsilon}, \\
 &V_{x,l}^{\sigma \sigma'}(\mathbf{k}) = -\frac{\delta_{\sigma \sigma'}}{A} \sum\limits_{k' \leq k_F} \frac{2\pi e^2}{\epsilon |\mathbf{k}-\mathbf{k}'|}, \\
&\Delta_\mathbf{k}^{\sigma \sigma'} = 0,
\end{split}
\end{equation}
where $f_\mathbf{k}$ is the Fermi-Dirac distribution.

The kinetic energy $E_{\rm kin}$, Hartree energy $E_{\rm H}$, exchange energy of the intralayer interaction $E_{x}^{\rm intra}$ and the interlayer interaction $E_{x}^{\rm inter}$ are, respectively,
\begin{equation}
\label{Eq_EtotS0}
\begin{split}
E_{\rm kin} &= 4 \sum\limits_{k \leq k_F} \frac{\hbar^2 k^2}{2m^*} = \frac{\pi \hbar^2 A}{4 m^*} n^2, \\
E_{\rm H} &= 0, \\
E_x^{\rm intra} &= -\frac{2}{A} \sum\limits_{k,k' \leq k_F} \frac{2\pi e^2}{\epsilon|\mathbf{k}-\mathbf{k}'|} = -\frac{4e^2 A}{3\sqrt{\pi} \epsilon} n^{3/2}, \\
E_x^{\rm inter} &= 0.
\end{split}
\end{equation}
We have used the analytical integrations over momenta $\mathbf{k}$ and $\mathbf{k}'$ summarized in Appendix~\ref{Appendix_integrals}.
The HF energy $E^{S_0}_{\rm tot}$ is the sum of these contributions in Eq.~(\ref{Eq_EtotS0}).
The HF energy per electron $\varepsilon^{S_0}_{\rm tot} = E^{S_0}_{\rm tot}/nA$, expressed in terms of $n$ and $r_s$, is
\begin{equation}
\begin{split}
\varepsilon^{S_0}_{\rm tot}(n)
&= \frac{\pi \hbar^2}{4m^*}n - \frac{4e^2}{3\sqrt{\pi} \epsilon} n^{1/2}, \\
\varepsilon^{S_0}_{\rm tot}(r_s)
&= \left[ \frac{1}{r_s^2} - \frac{8\sqrt{2}}{3\pi r_s} \right] \rm Ry^*.
\end{split}
\end{equation}
$r_sa^* = r_{s,t}a^* = r_{s,b}a^* = \sqrt{2/\pi n}$, here $r_{s,t}$ and $r_{s,b}$ are inter-electron distance in the top and bottom layers, respectively.

\subsubsection{HF energy of the spin polarized but pseudospin unpolarized state $S_1$}
\label{subsec_S1}
The wavefunction of the spin polarized but pseudospin unpolarized state $S_1$ is
\begin{equation}
\begin{split}
|S_1 \rangle &= \prod\limits_{k \leq k_F} c^\dagger_{\xi \uparrow \mathbf{k}} c^\dagger_{\bar{\xi} \uparrow \mathbf{k}} |0 \rangle \\
&= \prod\limits_{k \leq k_F} c^\dagger_{b \uparrow \mathbf{k}} c^\dagger_{t \uparrow \mathbf{k}} |0 \rangle.
\end{split}
\end{equation}
There are two equal Fermi surfaces with Fermi momentum $k_F = \sqrt{2\pi n}$. The densities and HF potentials of $S_1$ state are
\begin{equation}
\begin{split}
& n_t = n_b = \frac{n}{2},\\
& \rho_{ll'}^{\sigma \sigma'}(\mathbf{k}) = \delta_{ll'} \delta_{\sigma \sigma'} \delta_{\uparrow \sigma} f_\mathbf{k}, \\
& V_{H,t} = V_{H,b} = \frac{\pi e^2 dn}{\epsilon}, \\
& V_{x,l}^{\sigma \sigma'}(\mathbf{k}) = -\frac{\delta_{\sigma \sigma'} \delta_{\uparrow \sigma}}{A} \sum\limits_{k' \leq k_F} \frac{2\pi e^2}{\epsilon |\mathbf{k}-\mathbf{k}'|}, \\
& \Delta_\mathbf{k}^{\sigma \sigma'} = 0.
\end{split}
\end{equation}
We have assumed the majority spin to be $\sigma = \uparrow$.
The energies are
\begin{equation}
\label{Eq_EHF_S1}
\begin{split}
E_{\rm kin} &= 2 \sum\limits_{k \leq k_F} \frac{\hbar^2 k^2}{2m^*} = \frac{\pi \hbar^2 A}{2 m^*} n^2, \\
E_{\rm H} &= 0, \\
E_x^{\rm intra} &= -\frac{1}{A} \sum\limits_{k,k' \leq k_F} \frac{2\pi e^2}{\epsilon|\mathbf{k}-\mathbf{k}'|} = -\frac{4\sqrt{2} e^2 A}{3\sqrt{\pi} \epsilon} n^{3/2}, \\
E_x^{\rm inter} &= 0.
\end{split}
\end{equation}
The HF energy per electron
\begin{equation}
\label{Eq_HFE_S1}
\begin{split}
\varepsilon^{S_1}_{\rm tot}(n)
&= \frac{\pi \hbar^2}{2m^*}n - \frac{4\sqrt{2} e^2}{3\sqrt{\pi} \epsilon} n^{1/2}, \\
\varepsilon^{S_1}_{\rm tot}(r_s)
&= \left[ \frac{2}{r_s^2} - \frac{16}{3\pi r_s} \right] \rm Ry^*,
\end{split}
\end{equation}
and $r_sa^* = r_{s,t}a^* = r_{s,b}a^* = \sqrt{2/\pi n}$.

\subsubsection{HF energy of the spin polarized interlayer coherent state $S_\xi$}
The wavefunction of the spin polarized interlayer coherent (pseudospin polarized) state $S_\xi$ is
\begin{equation}
\begin{split}
|S_\xi \rangle &= \prod\limits_{k \leq k_F} c^\dagger_{\xi \uparrow \mathbf{k}} | 0\rangle \\
&= \prod\limits_{k \leq k_F} (\alpha c^\dagger_{t \uparrow \mathbf{k}} + \beta c^\dagger_{b \uparrow \mathbf{k}})| 0\rangle,
\end{split}
\end{equation}
The pseudospin is in a direction with polar angle $\theta = 2 \arctan(\alpha/\beta)$ with a freedom of any azimuthal angle $\phi$ even though we specifically choose $\alpha,\beta$ to be real.
There is only one Fermi surface with Fermi momentum $k_F = \sqrt{4\pi n}$ for the pseudospin fully polarized state. The densities and HF potentials of $S_\xi$ state are
\begin{equation}
\label{Eq_HFV_Sxi}
\begin{split}
& n_t = n \alpha^2, \quad n_b = n \beta^2, \\
& \rho_{ll'}^{\sigma \sigma'}(\mathbf{k}) = \delta_{\sigma \sigma'} \delta_{\uparrow \sigma} [\delta_{ll'} \delta_{lt} \alpha^2 + \delta_{ll'} \delta_{lb} \beta^2 + (1-\delta_{ll'}) \alpha\beta] f_\mathbf{k}, \\
& V_{H,l} = \frac{2\pi e^2 dn}{\epsilon} [\delta_{ll'} \delta_{lt} \alpha^2 + \delta_{ll'} \delta_{lb} \beta^2],  \\
& V_{x,l}^{\sigma \sigma'}(\mathbf{k}) = - [\delta_{ll'} \delta_{lt} \alpha^2 + \delta_{ll'} \delta_{lb} \beta^2] \frac{\delta_{\sigma \sigma'} \delta_{\uparrow \sigma}}{A}\sum\limits_{k' \leq k_{F}} \frac{2\pi e^2}{\epsilon |\mathbf{k}' - \mathbf{k}|}, \\
& \Delta^{\sigma \sigma'}_\mathbf{k} = \alpha \beta\frac{\delta_{\sigma \sigma'} \delta_{\uparrow \sigma}}{A} \sum\limits_{k' \leq k_F} \frac{2\pi e^2}{\epsilon |\mathbf{k}' - \mathbf{k}|} e^{-|\mathbf{k}'-\mathbf{k}|d}.
\end{split}
\end{equation}
The energies are
\begin{equation}
\label{Eq_SP_energies}
\begin{split}
E_{\rm kin} &= \sum\limits_{k \leq k_{F}} \frac{\hbar^2 k^2}{2m^*} 
= \frac{\pi \hbar^2A}{m^*} n^2, \\
E_{\rm H} &= \frac{1}{2} \sum\limits_{k \leq k_{F}} \frac{2\pi e^2 dn}{\epsilon} 
\frac{(\alpha^2 - \beta^2)^2}{2} \\
&= \frac{\pi e^2 d}{2\epsilon} n^2 A (\alpha^2 - \beta^2 )^2, \\
E^{\rm intra}_x
&= -\frac{1}{2A} \sum\limits_{k,k' \leq k_F}  \frac{2\pi e^2}{\epsilon |\mathbf{k}' - \mathbf{k}|} \Big( \alpha^4 + \beta^4 \Big) \\
&= -\frac{8e^2 A}{3\sqrt{\pi} \epsilon} n^{3/2} (\alpha^4+\beta^4), \\
E_x^{\rm inter}
&=-\frac{1}{2A} \sum\limits_{k,k' \leq k_F} \frac{2\pi e^2}{\epsilon |\mathbf{k} - \mathbf{k}'|} 2\alpha^2 \beta^2 e^{-|\mathbf{k} - \mathbf{k}'|d} \\
& = -\frac{2e^2A \alpha^2 \beta^2}{\sqrt{\pi} \epsilon} n^{3/2} J(k_Fd),
\end{split}
\end{equation}
where $J(k_Fd)$ is the triple integral
\begin{equation}
\label{Eq_JkFd}
\begin{split}
J(k_Fd) = \int_0^1 dx x \int_0^1 dy y &\int_0^{2\pi} d\theta \\
&\frac{e^{-k_Fd \sqrt{x^2+y^2-2xy \cos \theta}}}{\sqrt{x^2+y^2-2xy \cos \theta}},
\end{split}
\end{equation}
which should be evaluated numerically.
The HF energy per electron is
\begin{equation}
\label{Eq_SP_Etot1}
\begin{split}
\varepsilon_{\rm tot}^{S_\xi}(n)
&= \frac{\pi \hbar^2}{m^*} n +  \frac{\pi e^2d}{2\epsilon} n (\alpha^2 - \beta^2 )^2 \\
& - \frac{8e^2}{3\sqrt{\pi} \epsilon}  n^{1/2} 
\left( \alpha^4+\beta^4 + \frac{3}{4}J(k_Fd) \alpha^2 \beta^2 \right).
\end{split}
\end{equation}
Represented in dimensionless length using
$r_{s,l} a^* = 1/\sqrt{\pi n_l}$ and
\begin{equation}
\begin{split}
\pi (n_t + n_b) &= \pi n \\
&= \Big( \frac{1}{r_{s,t}^2} + \frac{1}{r_{s,b}^2} \Big) \frac{1}{(a^*)^2} \\
&\equiv \frac{1}{(\tilde{r}_s a^*)^2},
\end{split}
\end{equation}
the HF energy per electron is
\begin{equation}
\label{Eq_SP_Etot2}
\begin{split}
\varepsilon^{S_\xi}_{\rm tot} (\tilde{r}_s)
&= \Big[ \frac{2}{\tilde{r}_s^2} + \frac{d}{a^* \tilde{r}_s^2}(\alpha^2-\beta^2)^2 \\
&- \frac{16}{3\pi \tilde{r}_s} \Big( \alpha^4+\beta^4 + \frac{3}{4}J(k_Fd) \alpha^2 \beta^2 \Big) \Big]
{\rm Ry}^*.
\end{split}
\end{equation}
Note that $\tilde{r}_s \equiv \tilde{r}_s(r_{s,t}, r_{s,b})$.

In particular, the $S_3$ phase, the spin polarized interlayer coherent state with pseudospin polarized in the $z$-direction (polar angle $\theta=0$), is the case that all electrons are in one layer but not the other, i.e., $\alpha = 1$, $\beta=0$, $\tilde{r}_s = r_{s,t}$ and $r_{s,b} \rightarrow \infty$. The HF energy of $S_3$ phase is
\begin{equation}
\label{Eq_Etot_S3}
\begin{split}
&\varepsilon^{S_3}_{\rm tot}(n) = \frac{\pi \hbar^2}{m^*} n +  \frac{\pi e^2}{2\epsilon} nd 
- \frac{8e^2}{3\sqrt{\pi} \epsilon}  n^{1/2}, \\
&\varepsilon^{S_3}_{\rm tot}(\tilde{r}_s) = \Big[ \frac{2}{\tilde{r}_s^2} +  \frac{d}{a^* \tilde{r}^2_s}
- \frac{16}{3\pi \tilde{r}_s} \Big] {\rm Ry}^*.
\end{split}
\end{equation}

The $S_2$ phase, the spin polarized interlayer coherent state with pseudospin polarized in the $xy$-plane (polar angle $\theta=\pi/2$), is the case of equal layer densities, i.e., $\alpha^2 = \beta^2 = 1/2$, $r_s = r_{s,t} = r_{s,b} = \sqrt{2} \tilde{r}_s$. The HF energy per electron of $S_2$ phase is
\begin{equation}
\label{Eq_Etot_S2}
\begin{split}
\varepsilon^{S_2}_{\rm tot}(n) &= \frac{\pi \hbar^2}{m^*} n - \frac{4e^2}{3\sqrt{\pi} \epsilon}  n^{1/2} 
\left( 1 + \frac{3}{8}J(k_Fd) \right), \\
\varepsilon^{S_2}_{\rm tot}(r_s)
&= \Big[\frac{4}{r_s^2} 
- \frac{8\sqrt{2}}{3\pi r_s}
\Big(1 + \frac{3}{8} J(k_Fd) \Big)
\Big]{\rm Ry^*}, \\
\varepsilon^{S_2}_{\rm tot}(\tilde{r}_s)
&= \Big[\frac{2}{\tilde{r}_s^2} 
- \frac{8}{3\pi \tilde{r}_s}
\Big(1 + \frac{3}{8} J(k_Fd) \Big)
\Big]{\rm Ry^*}.
\end{split}
\end{equation}

The triple integral $J(k_Fd)$ defined in Eq.~(\ref{Eq_JkFd}) has the properties that 
\begin{equation}
\begin{cases}
& 0 \leq J(k_Fd) \leq \frac{8}{3}, \\
&J(0) = \frac{8}{3}, \\
&\lim\limits_{k_Fd \gg 1} J(k_Fd) \rightarrow 0.
\end{cases}
\end{equation}
As expected, for $d=0$, 
\begin{equation}
\begin{split}
&\lim\limits_{d \rightarrow 0} \varepsilon^{S_2}_{\rm tot}(n) \rightarrow \frac{\pi \hbar^2}{m^*} n - \frac{8e^2}{3\sqrt{\pi} \epsilon}  n^{1/2}, \\
&\lim\limits_{d \rightarrow 0} \varepsilon^{S_2}_{\rm tot}(r_s) \rightarrow \Big[\frac{4}{r_s^2} - \frac{16\sqrt{2}}{3\pi r_s} \Big] {\rm Ry}^*,
\end{split}
\end{equation}
which recovers the energy of single 2DEG with one Fermi surface.

For $0 < k_Fd \ll 1$, $J(k_Fd)$ can be expanded in Taylor series 
\begin{widetext}
\begin{equation}
\label{Eq_Jd_expand}
\begin{split}
J(k_Fd) &= J(0) - \frac{\pi}{2} k_Fd + \frac{1}{2} (k_F d)^2 \int_0^1 dx x \int_0^1 dy y \int_0^{2\pi} d\theta \sqrt{x^2+y^2-2xy\cos\theta} + \cdots 
\end{split}
\end{equation}
\end{widetext}

\subsection{The phase diagram for equal layer densities}
\label{subsec_phase_equaln}
The $S_2$ phase with XY easy plane pseudospin ordering always has a lower energy than the Ising-ordered $S_3$ phase (with all electrons in one layer with pseudospin polarized along z direction) because the density imbalance between the layers in $S_3$ pays an extra Hartree energy. 
This preference is also inferred from Eq.~(\ref{Eq_Etot_S3}) and Eq.~(\ref{Eq_Etot_S2})
by taking their energy difference:
\begin{equation}
\begin{split}
&\varepsilon^{S_2}_{\rm tot}(n) - \varepsilon^{S_3}_{\rm tot}(n) \\
&= -\frac{4e^2}{3\sqrt{\pi} \epsilon} n^{1/2} \Big(\frac{3}{8}J(k_Fd)-1 \Big) - \frac{\pi e^2}{2\epsilon} nd \\
&= -\frac{e^2k_F}{4 \pi \epsilon} \Big[ J(k_Fd)- \frac{8}{3} + \frac{\pi}{2} k_Fd \Big].
\end{split}
\end{equation}
The function $J(z)$ monotonically decreases with $z$ and intersects with the line $-\pi z/2 + 8/3$ at $z= 0$, the slope of which equals the derivative of $J(z)$ at $z = 0$,
\begin{equation}
\frac{d J(z)}{dz} \Big|_{z=0} = -\frac{\pi}{2}.
\end{equation}
It follows that
\begin{equation}
\varepsilon^{S_2}_{\rm tot} - \varepsilon^{S_3}_{\rm tot} < 0 \quad \forall \  k_Fd > 0.
\end{equation}
This energetic hierarchy is shown by the computed HF energies as a function of layer separation $d$ in Appendix~\ref{HF_energy_T0}.

\begin{figure}
\centering
\includegraphics[width=0.85\columnwidth]{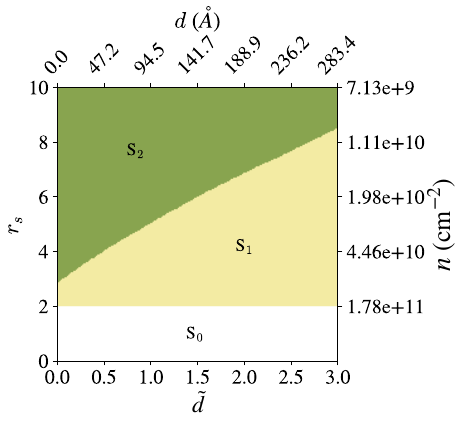}
\caption{\label{fig_phase_rsd} {
 The zero-temperature phase diagram with respect to $r_s$ and $\tilde{d}$, where $\tilde{d} = d/a^*$. Dimensionless $r_s$ and $\tilde{d}$ are also converted to density $n$ in the unit of cm$^{-2}$ and distance $d$ in the unit of $\text{\AA}$ using GaAs quantum well parameters. $S_0$, $S_1$ and $S_2$ phases all have equal layer densities, i.e., $r_sa^* = r_{st}a^* = r_{sb}a^* = \sqrt{2/\pi n}$.
  }}
\end{figure}

The zero-temperature phase diagram, determined by $\varepsilon^{G}_{\rm tot} = \min\{ \varepsilon^{S_0}_{\rm tot}, \varepsilon^{S_1}_{\rm tot}, \varepsilon^{S_2}_{\rm tot}\}$, with respect to $r_s$ and $\tilde{d}$ is shown in Fig.~\ref{fig_phase_rsd}. 
The interlayer coherent phase $S_2$ manifests stability at larger $r_s$ (lower electron density) and smaller interlayer distance $\tilde{d}$.
Note that there is always a critical $r_s$ for the transition to the interlayer coherent phase, which is stable only above a specific $d$-dependent $r_s$ value, with the critical $r_s$ increasing with increasing $d$.

\subsection{The phase diagram for unequal layer densities}
\label{subsec_unequaln}
When the two layers have unequal densities, the spin polarization of each layer individually depends on the layer density: if the layer density is larger (smaller) than the critical value corresponding to $r_{s} \sim 2$, the spin-unpolarized (spin-polarized) phase is favored.
The phase that the lower-density layer is spin-polarized and the higher-density layer is spin-unpolarized has been identified in a previous theoretical study\cite{Hanna_bilayer_2000} as the three-component phase.
In the rest of the paper, we assume the density in each layer is lower than this critical value of the ferromagnetic instability ($r_s \sim 2$) and therefore will only consider the spin polarized case because spin plays no role in the interlayer coherence physics.
For unequal layer densities, we compare HF energies of the interlayer coherent phase $S_\xi$ in Eq.~(\ref{Eq_SP_Etot1}) with the interlayer incoherent phase $S_1'$. The $S_1'$ phase is the generalized case of $S_1$ phase (defined in Sec.~\ref{subsec_S1}) but with two unequal Fermi surfaces. It is straightforward to write down the HF energy per electron of $S_1'$ phase, represented in total density $n$ and layer polarization $m=(n_t-n_b)/n$:
\begin{gather}
\varepsilon^{S_1'}_{\rm tot}(n,m)
= \frac{\pi \hbar^2}{2m^*}n (1+m^2) 
+ \frac{\pi e^2d}{2\epsilon}nm^2 \nonumber\\
\qquad - \frac{2\sqrt{2} e^2}{3\sqrt{\pi} \epsilon} n^{1/2} \Big[(1+m)^{3/2} + (1-m)^{3/2} \Big].
\end{gather}
For a direct comparison, we represent the HF energy of $S_\xi$ phase in Eq.~(\ref{Eq_SP_Etot1}) using $n$ and $m$ as well:
\begin{equation}
\label{Eq_Etot_Sxi}
\begin{split}
\varepsilon_{\rm tot}^{S_\xi}(n,m)
&= \frac{\pi \hbar^2}{m^*} n +  \frac{\pi e^2d}{2\epsilon} n m^2 \\
& - \frac{4e^2}{3\sqrt{\pi} \epsilon}  n^{1/2} 
\Big[ 1+ m^2 + \frac{3}{8}J(k_Fd) (1- m^2)\Big].
\end{split}
\end{equation}
For layer unpolarized case $m=0$, $S_1'$ phase is equivalent to the pseudospin unpolarized phase $S_1$, $\varepsilon^{S_1'}_{\rm tot}(n,m=0) = \varepsilon^{S_1}_{\rm tot}(n)$. For the totally layer polarized case $m=1$, $S_1'$ phase is equivalent to $S_\xi$, there is only one Fermi surface and no interlayer coherence because one layer is completely empty.

\begin{figure*}[!htb]
\centering
\includegraphics[width=1.0\textwidth]{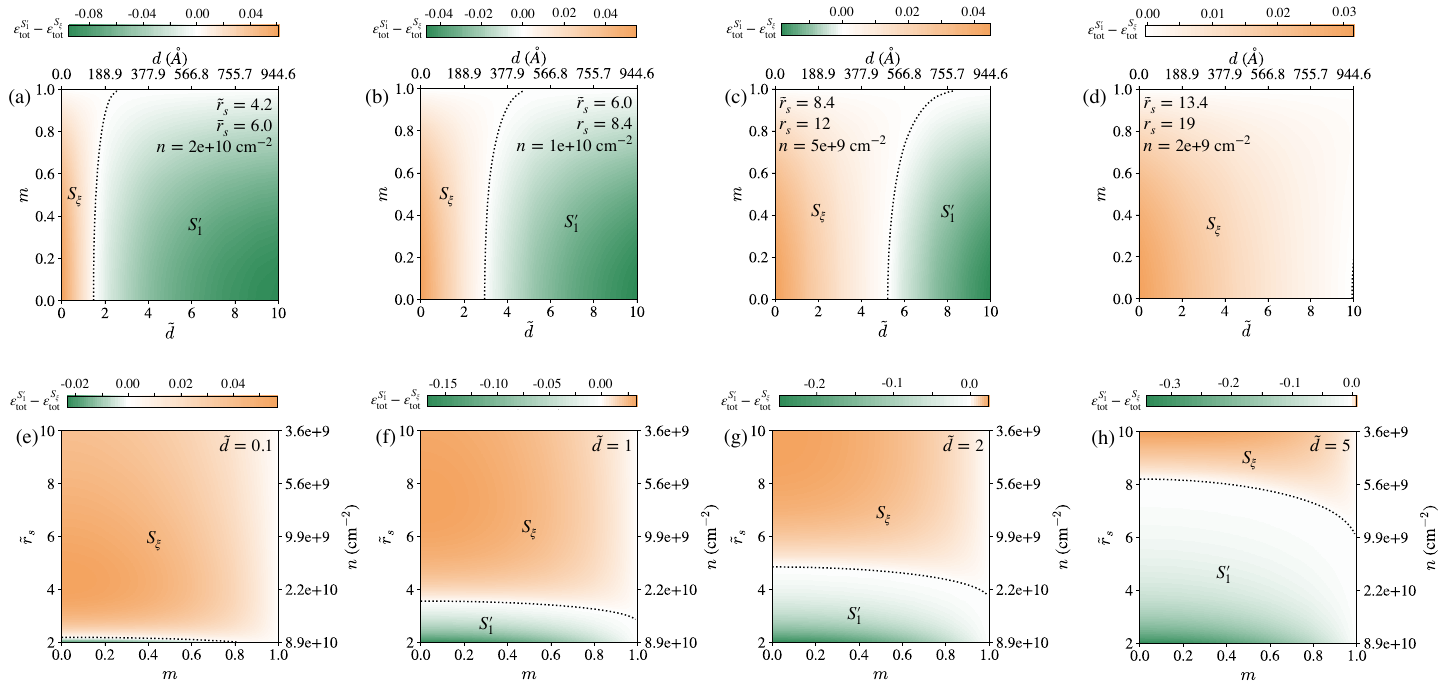}
\caption{\label{fig_Ediff_S1p_SP} {
Energy difference $\varepsilon_{\rm tot}^{S_1'} - \varepsilon_{\rm tot}^{S_\xi}$. (a-d) As a function of layer polarization $m$ and $\tilde{d}$ for fixed total densities $n=2 \times 10^{10}$ cm$^{-2}$, $10^{10}$ cm$^{-2}$, $5 \times 10^{9}$ cm$^{-2}$ and $2 \times 10^{9}$ cm$^{-2}$.
 The corresponding $\tilde{r}_s$ (defined by the total density, $\tilde{r}_s a^*=\sqrt{1/\pi n}$) and $\bar{r}_s$ (defined by the averaged layer density, $\bar{r}_s a^* = \sqrt{2/\pi n}$) are labeled in each subplot.
(e-h) As a function of $\tilde{r}_s$ and $m$ for fixed $\tilde{d}=0.1, 1, 2$ and $5$. The dotted lines trace the phase boundaries.
As $m$ increases, the phase boundary tilts to larger $\tilde{d}$ and smaller $\tilde{r}_s$, indicating that the interlayer coherent phase $S_\xi$ is more preferred for larger $m$.
  }}
\end{figure*}

In Fig.~\ref{fig_Ediff_S1p_SP}(a-d) we plot the energy difference $\varepsilon_{\rm tot}^{S_1'} - \varepsilon_{\rm tot}^{S_\xi}$ as a function of layer polarization $m$ and interlayer separation $\tilde{d}$ for several fixed total densities. The corresponding $\tilde{r}_s$ (defined by the total density, $\tilde{r}_s a^*=\sqrt{1/\pi n}$) and $\bar{r}_s$ (defined by the average layer density, $\bar{r}_s a^* = \sqrt{2/\pi n}$) are labeled in each subplot. 
Similarly, Fig.~\ref{fig_Ediff_S1p_SP}(e-h) plot $\varepsilon_{\rm tot}^{S_1'} - \varepsilon_{\rm tot}^{S_\xi}$ as a function of $\tilde{r}_s$ and $m$ for several fixed layer separations. 
As $m$ increases, the phase boundary shifts to larger $\tilde{d}$ (Fig.~\ref{fig_Ediff_S1p_SP}(a-d)) and smaller $\tilde{r}_s$ (Fig.~\ref{fig_Ediff_S1p_SP}(e-h)), indicating that the interlayer coherent phase $S_\xi$ is more preferred than the interlayer incoherent phase $S_1'$ for larger $m$.
We conclude from Fig.~\ref{fig_Ediff_S1p_SP} that the interlayer coherent phase $S_\xi$ is preferred for all $m \in [0,1)$.

For the two interlayer incoherent phases $S_1$ and $S_1'$, their energy difference depends on layer polarization $m$ by
\begin{align}
\varepsilon^{S_1'}_{\rm tot} - \varepsilon^{S_1}_{\rm tot} 
&= \frac{\pi \hbar^2}{2m^*}nm^2 
+ \frac{\pi e^2d}{2\epsilon}nm^2 \nonumber\\
 - \frac{2\sqrt{2} e^2}{3\sqrt{\pi} \epsilon} &n^{1/2} \Big[(1+m)^{3/2} + (1-m)^{3/2} -2 \Big].
\end{align}
Expand in $m$,
\begin{align}
\frac{\varepsilon^{S_1'}_{\rm tot} - \varepsilon^{S_1}_{\rm tot}}{\text{Ry}^*}
= &\pi a^* m^2 \sqrt{n} \big[ (a^*+d) \sqrt{n} - \frac{\sqrt{2}}{\pi^{3/2}} \big] \nonumber\\
&+ \mathcal{O}(m^4),
\end{align}
$\varepsilon^{S_1'}_{\rm tot} - \varepsilon^{S_1}_{\rm tot} < 0$ only for small $n$ and small $d$.

\begin{figure*}[!htb]
\centering
\includegraphics[width=1.0\textwidth]{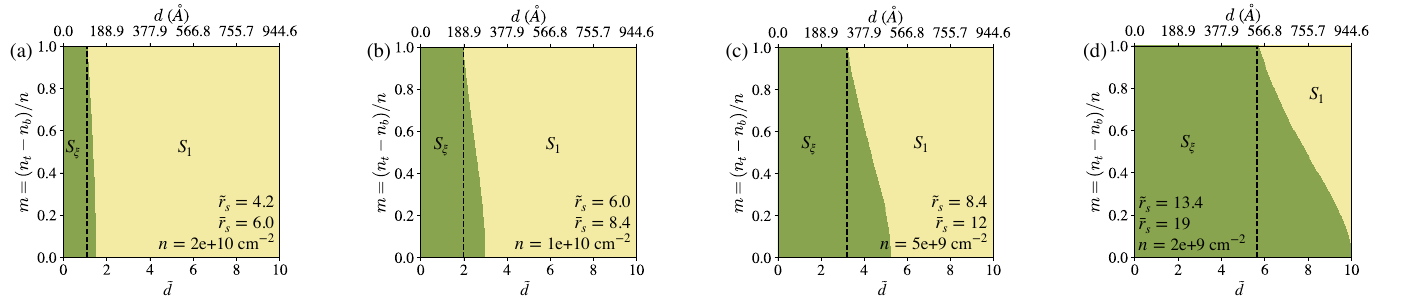}
\caption{\label{fig_phase_m-d} {
 Phase diagrams as a function of layer polarization $m$ and $\tilde{d}$ for fixed total densities $n=2 \times 10^{10}$ cm$^{-2}$, $10^{10}$ cm$^{-2}$, $5 \times 10^{9}$ cm$^{-2}$ and $2 \times 10^{9}$ cm$^{-2}$.
 The phase diagrams are determined by finding the lowest energy $\min\{\varepsilon_{\rm tot}^{S_1}, \varepsilon_{\rm tot}^{S'_1}, \varepsilon_{\rm tot}^{S_\xi}\}$.
 The corresponding $\tilde{r}_s$ (defined by the total density, $\tilde{r}_s a^*=\sqrt{1/\pi n}$) and $\bar{r}_s$ (defined by the averaged layer density, $\bar{r}_s a^* = \sqrt{2/\pi n}$) are labeled in each subplot.
 The black dashed lines plot the critical distance $\tilde{d}_c$ in Eq.~(\ref{Eq_dc}), converging precisely to the phase transition boundary for $m \rightarrow 1$ and for $n \rightarrow n_{\rm max}$ as estimated in Eq.~(\ref{Eq_nmax}).
  }}
\end{figure*}

In Fig.~\ref{fig_phase_m-d}, we show phase diagrams as a function of $m$ and $\tilde{d}$ for fixed total densities, chosen to be the same as in Fig.~\ref{fig_Ediff_S1p_SP}(a-d).
The phase diagrams are determined by the lowest energy $\min\{\varepsilon_{\rm tot}^{S_1}, \varepsilon_{\rm tot}^{S'_1}, \varepsilon_{\rm tot}^{S_\xi}\}$. 
In these phase diagrams, the $S_1'$ phase is never the ground state, even for larger $\tilde{d}$ where $S_1'$ is lower in energy than $S_\xi$ as in Fig.~\ref{fig_Ediff_S1p_SP}(a-d). This is because the critical $\tilde{d}$ for $\varepsilon^{S_1'}_{\rm tot} - \varepsilon^{S_1}_{\rm tot} < 0$ is smaller than that of the phase boundary in Fig.~\ref{fig_Ediff_S1p_SP}(a-d).

For $m \rightarrow 1$, the phase boundary in Fig.~\ref{fig_phase_m-d} can be understood analytically by examining the energy difference between the interlayer coherent phase $S_\xi$ and incoherent phase $S_1$:
\begin{widetext}
\begin{equation}
\label{Eq_Ediff}
\begin{split}
\frac{\varepsilon_{\rm tot}^{S_\xi} - \varepsilon_{\rm tot}^{S_1}}{\text{Ry}^*} &= \pi n a^* [ a^* + d(\alpha^2-\beta^2)^2] + \frac{n^{1/2}a^*}{\pi^{1/2}} \Big[ 4\alpha^2 \beta^2 (\frac{8}{3}-J(k_Fd)) - \frac{8(2-\sqrt{2})}{3} \Big] \\
&= \pi n a^* [ a^* + dm^2] + \frac{n^{1/2}a^*}{\pi^{1/2}} \Big[ (1-m^2) (\frac{8}{3}-J(k_Fd)) - \frac{8(2-\sqrt{2})}{3} \Big] .
\end{split}
\end{equation}
\end{widetext}
When $m \rightarrow 1$,
\begin{equation}
\lim\limits_{m \rightarrow 1} \frac{\varepsilon_{\rm tot}^{S_\xi} - \varepsilon_{\rm tot}^{S_1}}{\text{Ry}^*}= \pi n a^*(a^* + d) - \frac{n^{1/2}a^*}{\pi^{1/2}}  \frac{8(2-\sqrt{2})}{3} ,
\end{equation}
setting it to zero, we find that 
\begin{equation}
\label{Eq_dc_m1}
\begin{split}
\tilde{d}_c\big|_{m= 1} &\equiv \frac{d_c}{a^*} \Big|_{m=1} \\
&= \frac{8(2-\sqrt{2})}{3} \frac{1}{\pi^{3/2} n^{1/2}a^*} - 1,
\end{split}
\end{equation}
which only depends on total density $n$. The $\tilde{d}_c$ in Eq.~(\ref{Eq_dc_m1}) is shown in black dashed lines in Fig.~\ref{fig_phase_m-d}, they agree well with the phase boundary when $m \rightarrow 1$.

In fact, if we take a close look at the second line of Eq.~(\ref{Eq_Ediff}), setting it to zero gives
\begin{equation}
\label{Eq_condition}
\begin{split}
\pi^{3/2} &n^{1/2}(a^* + d_c m^2) \\
&+ \Big[(1-m^2)(\frac{8}{3}-J(k_Fd_c)) - \frac{8(2-\sqrt{2})}{3} \Big] = 0.
\end{split}
\end{equation}
The first term $\pi^{3/2} n^{1/2} (a^*+d_cm^2)$ is positive, the only way that a $d_c$ exists is that the second term is negative, which requires
\begin{equation}
\label{Eq_Jcondition}
J(k_Fd_c) > \frac{8}{3} - \frac{8(2-\sqrt{2})}{3(1-m^2)}.
\end{equation}
Given that $J(k_Fd_c) \in [0, 8/3]$ and it decays rapidly with $k_Fd_c$, i.e., with $n^{1/2}d_c$, the competition between the first and the second term in Eq.~(\ref{Eq_condition}) in solving for a $d_c$, if it exists, ultimately requires $k_F d_c$ to be small for large $n$, and validates the approximation
\begin{equation}
\label{Eq_J_approx}
J(k_Fd) \approx \frac{8}{3} - \frac{\pi}{2}k_Fd,
\end{equation}
which are the first two terms in the expansion of Eq.~(\ref{Eq_Jd_expand}).
Substitute Eq.~(\ref{Eq_J_approx}) to Eq.~(\ref{Eq_condition}), we have
\begin{equation}
\begin{split}
\tilde{d}_c \big|_{n \rightarrow n_{\rm max}} = \frac{8(2-\sqrt{2})}{3}\frac{1}{\pi^{3/2} n^{1/2} a^*} - 1.
\end{split}
\end{equation}
The $m$ dependence of $d_c$ vanishes for large densities, and this critical distance $d_c$ equals the one for $m \rightarrow 1$ case in Eq.~(\ref{Eq_dc_m1}).
Taking $m \rightarrow 0$, Eq.~(\ref{Eq_condition}) becomes
\begin{equation}
J(k_Fd_c) = \pi^{3/2} n^{1/2} a^* + \frac{8(\sqrt{2}-1)}{3}.
\end{equation}
Because $J(k_Fd_c) \leq 8/3$, we could find that the maximum density for a $d_c$ to exist is
\begin{equation}
\label{Eq_nmax}
n_{\rm max} \sim 8.8 \times 10^{10} {\rm cm}^{-2}.
\end{equation}

From the discussion above, we find that the critical distance $\tilde{d}_c$ at these two limits, $m \rightarrow 1$ and $n \rightarrow n_{\rm max}$, turns out to be the same value
\begin{equation}
\label{Eq_dc}
\begin{split}
\tilde{d}_c \big|_{m= 1} &= \tilde{d}_c \big|_{n \rightarrow n_{\rm max}} \\
&= \frac{8(2-\sqrt{2})}{3}\frac{1}{\pi^{3/2} n^{1/2} a^*} - 1.
\end{split}
\end{equation}
This explains the agreement between the phase boundary and the black dashed lines in Fig.~\ref{fig_phase_m-d}, when either the layer density imbalance is large, as in Fig.~\ref{fig_phase_m-d}(a-d) at $m \rightarrow 1$, or the total density is close to $n_{\rm max}$, as in Fig.~\ref{fig_phase_m-d}(a).

In Fig.~\ref{fig_phase_m-d}, the phase transition boundary $\tilde{d}_c$ decreases with layer polarization $m$, revealing that a layer density imbalance suppresses the interlayer coherent phase $S_\xi$. This suppression becomes weaker as total density $n$ increases.
When $n$ approaches $n_{\rm max}$, $\tilde{d}_c$ becomes almost independent of $m$. As long as $n$ remains below $n_{\rm max}$, a $\tilde{d}_c$ exists and the interlayer coherent phase survives even when the layer is almost completely polarized.

The HF energy of the interlayer coherent phase $S_\xi$ monotonically increases with layer polarization $m$. 
As seen by taking the derivative of Eq.~(\ref{Eq_Etot_Sxi}) with respect to $m$:
\begin{equation}
\begin{split}
\frac{d\varepsilon_{\rm tot}^{S_\xi}}{dm}
&= \frac{e^2 n^{1/2} m}{\epsilon \sqrt{\pi}} \Big( \pi^{3/2} n^{1/2} d + J(k_Fd) - \frac{8}{3} \Big),
\end{split}
\end{equation}
the term in the bracket is always positive for $d>0$, which is clear from the Taylor expansion in Eq.~(\ref{Eq_Jd_expand}). 
Therefore,
\begin{equation}
\frac{d\varepsilon_{\rm tot}^{S_\xi}}{dm} \geq 0, \quad \forall \ n,
\end{equation}
$\varepsilon_{\rm tot}^{S_\xi}$ monotonically increases with layer polarization $m$ for any density $n$.
Among interlayer coherent phases $S_\xi$, the layer fully polarized phase $S_3$ always has the highest energy and the layer equally occupied phase $S_2$ has the lowest energy.
If carriers in the two layers are allowed to transfer, the system is always stablized to the pseudospin XY-ordered phase $S_2$.

The layer polarization $m$ can be interpreted as the pseudospin response to an effective pseudospin magnetic field applied in the $z$-direction, which can be experimentally tuned by gate voltages. 
The bilayer system is trivially pseudospin-polarized in the $z$-direction by virtue of 
this effective magnetic field, which is zero for $m=0$ and fully polarized in $z$-direction for $m=1$.
Figure~\ref{fig_phase_m-d} illustrates that even when there is a strong effective pseudospin magnetic field polarizing the pseudospin completely in the $z$-direction, the exchange-driven XY pseudospin ferromagnetic transition is little affected.

\begin{figure*}
\centering
\includegraphics[width=1.0\textwidth]{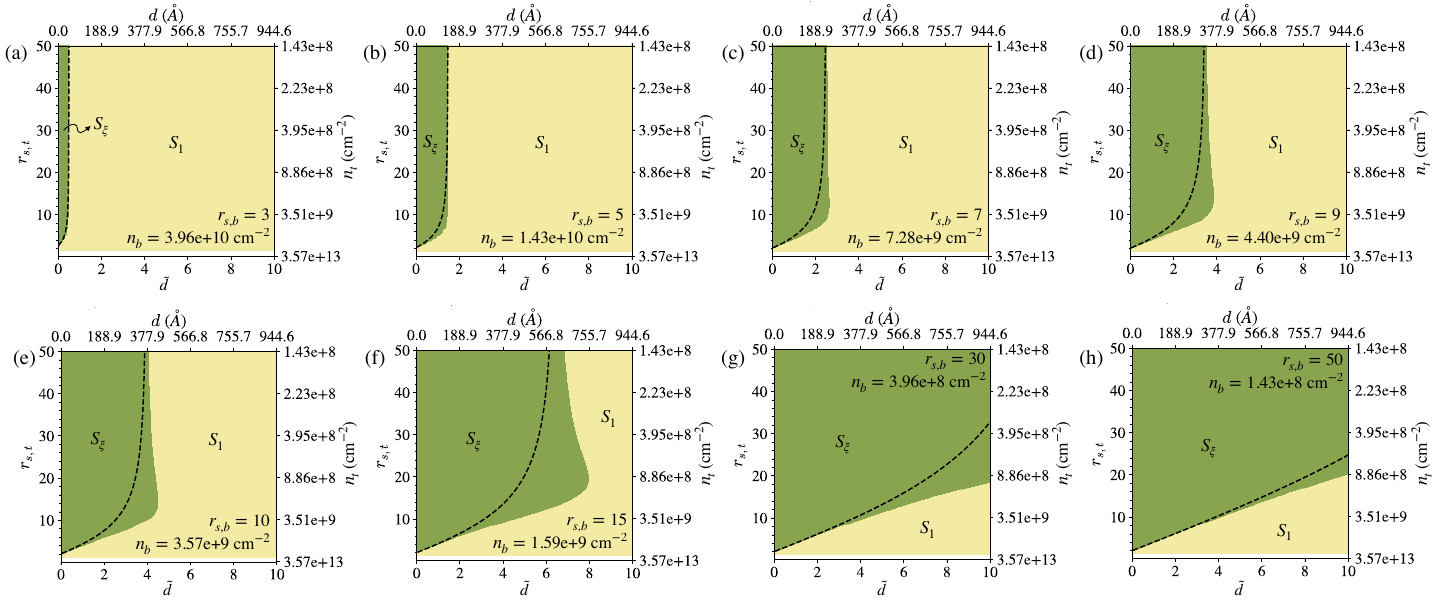}
\caption{\label{fig_phase_rs1-d} {
 Phase diagrams with respect to ($r_{s,t}$, $\tilde{d}$) for fixed $r_{s,b} \in [3, 50]$, corresponding to $n_b \sim\in [4 \times 10^{10}, 1.4 \times 10^{8}]$ cm$^{-2}$ in GaAs quantum wells.
 The black dashed lines plot the critical separation $\tilde{d}_c$ in Eq.~(\ref{Eq_dc}), which agrees well with the phase boundary when either the layer density imbalance is large or the total density is large: $\tilde{d}_c$ is linear in $r_{s,t}$ for $r_{s,t} \ll r_{s,b}$ and independent of $r_{s,t}$ for $r_{s,b} \ll r_{s,t}$.
 Note that for $r_{s,t} \lesssim 2$, the top layer is spin-unpolarized and the bottom layer is spin-polarized (because $r_{s,b} > 2$ in all presented figures) in the ground state.\cite{Hanna_bilayer_2000}}
  }
\end{figure*}

In Fig.~\ref{fig_phase_rs1-d}, we fix $r_{s,b}$ and evaluate the phase diagram as a function of $(r_{s,t},\tilde{d})$.
Note that for $r_{s,t} \lesssim 2$ in Fig.~\ref{fig_phase_rs1-d}, the top layer is spin-unpolarized and the bottom layer is spin-polarized (because $r_{s,b} > 2$ in all presented figures) in the ground state.\cite{Hanna_bilayer_2000}
In $m \rightarrow 1$ limit, $n_t \rightarrow n$, $n_b \rightarrow 0$, and $r_{s,t} \ll r_{s,b}$,
\begin{equation}
\tilde{d}_c \big|_{m=1} = \frac{8(2-\sqrt{2})}{3} \frac{r_{s,t}}{\pi} -1
\end{equation}
which is linear in $r_{s,t}$. In $m \rightarrow -1$ limit, $n_t \rightarrow 0$, $n_b \rightarrow n$, and $r_{s,b} \ll r_{s,t}$,
\begin{equation}
\tilde{d}_c \big|_{m=-1} = \frac{8(2-\sqrt{2})}{3} \frac{r_{s,b}}{\pi} -1
\end{equation}
which is independent of $r_{s,t}$.
In Fig.~\ref{fig_phase_rs1-d}, we plot $\tilde{d}_c$ in Eq.~(\ref{Eq_dc_m1}) in black dashed lines and it agrees well with the phase boundary when the density imbalance is large: linear in $r_{s,t}$ for $r_{s,t} \ll r_{s,b}$ and independent of $r_{s,t}$ for $r_{s,b} \ll r_{s,t}$.


\begin{figure*}[!htb]
\centering
\includegraphics[width=1.0\textwidth]{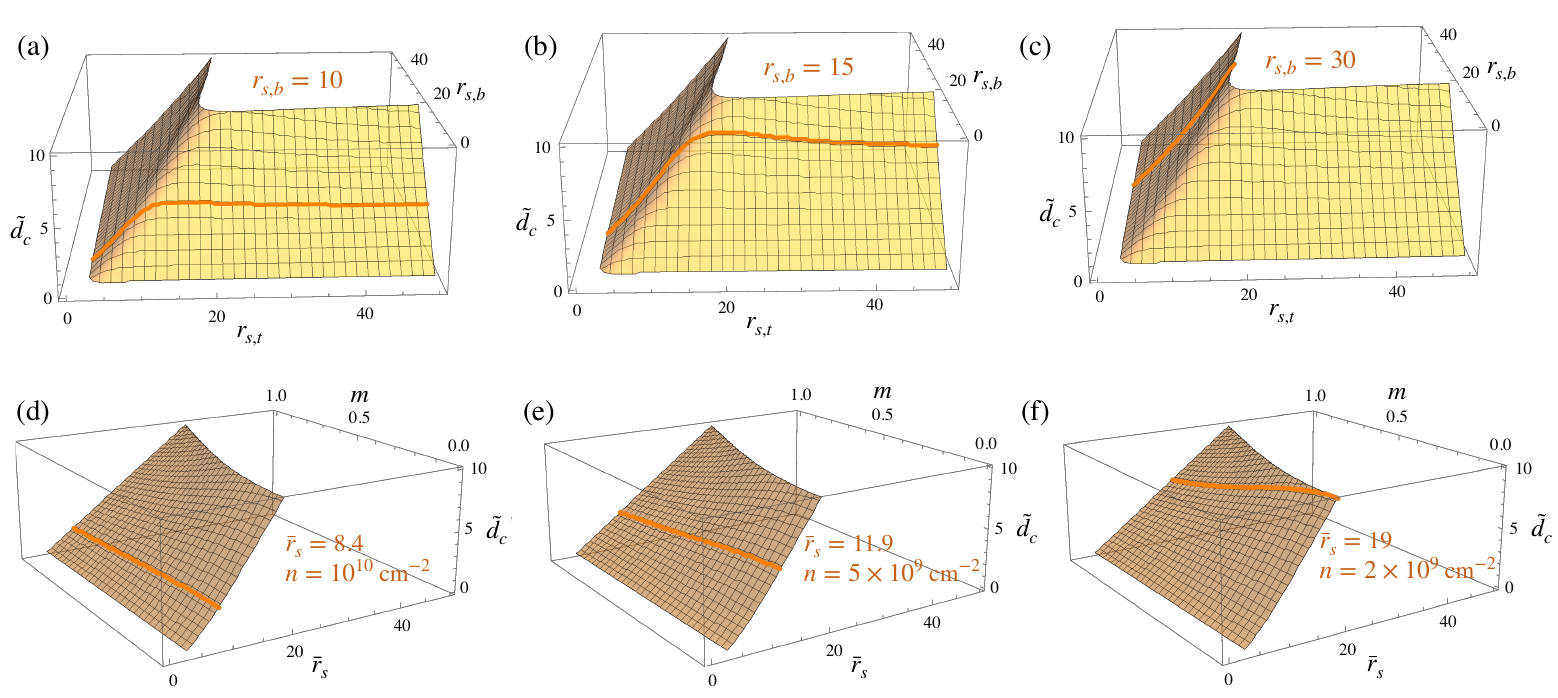}
\caption{\label{fig_phase3d} {
3D plots of the critical layer separation $\tilde{d}_c$. (a-c) With respect to ($r_{s,t}$,$r_{s,b}$). The orange linecuts trace the phase boundary $\tilde{d}_c$ versus $r_{s,t}$ at $r_{s,b}=10$, $15$ and $30$. These linecuts can be directly compared with the phase boundary in Fig.~\ref{fig_phase_rs1-d}(e-g).
(d-f) With respect to ($\bar{r}_{s}$,$m$), where $\bar{r}_s$ is defined by the average layer density using 
$\bar{r}_s a^* = \sqrt{2/\pi n}$. The orange linecuts trace the phase boundary $\tilde{d}_c$ versus $m$ at $\bar{r}_{s}=8.4$, $11.9$ and $26.7$. These linecuts can be directly compared with the phase boundary in Fig.~\ref{fig_phase_m-d}(b-d).
  }}
\end{figure*}

To provide a comprehensive picture of the phase diagram, we present 3D plots of the critical distance $\tilde{d}_c$ with respect to $(r_{s,t}, r_{s,b})$ in Fig.~\ref{fig_phase3d}(a-c) and with respect to $(\bar{r}_{s}, m)$ in Fig.~\ref{fig_phase3d}(d-f). 
Particularly, we mark the linecuts of the phase boundary, i.e. $\tilde{d}_c$ versus $r_{s,t}$ at $r_{s,b}=10, 15$ and $30$ in Fig.~\ref{fig_phase3d}(a-c), which can be directly compared with the phase transition boundary in 
Fig.~\ref{fig_phase_rs1-d}(e-g). Similarly, we show the linecuts of the phase boundary $\tilde{d}_c$ versus $m$
at $\bar{r}_s=8.4, 11.9$ and $26.7$ in Fig.~\ref{fig_phase3d}(d-f), as a direct comparison with the phase transition boundary in 
Fig.~\ref{fig_phase_m-d}(b-d).

\begin{figure}
\centering
\includegraphics[width=1.0\columnwidth]{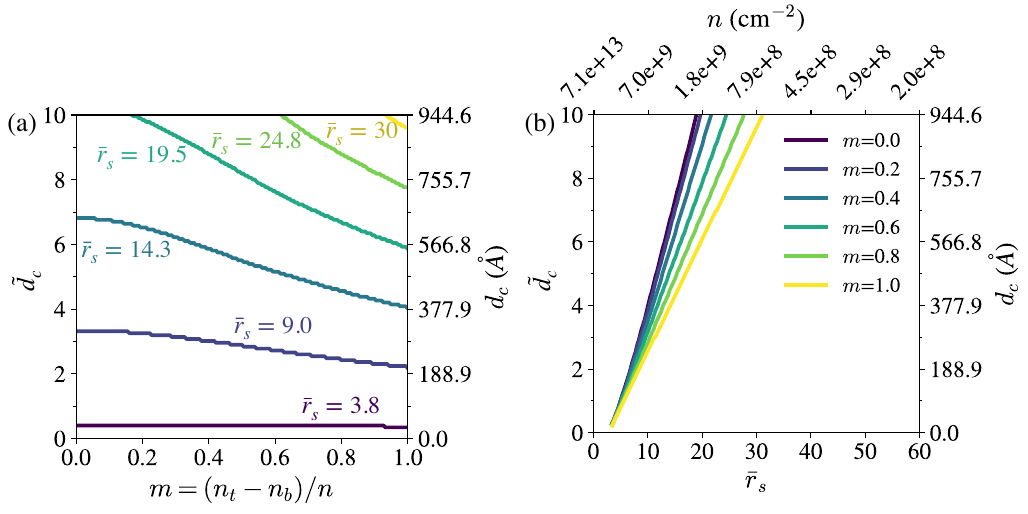}
\caption{\label{fig_dc_vs_X} {
(a) $\tilde{d}_c$ versus layer polarization $m$ for six values of $\bar{r}_s \in [3.8,30]$.
(b) $\tilde{d}_c$ versus $\bar{r}_s$ for six values of $m \in [0,1]$.
$\bar{r}_s$ is defined by the average layer density using $\bar{r}_s a^* = \sqrt{2/\pi n}$.
  }}
\end{figure}

\begin{figure*}
\centering
\includegraphics[width=1.0\textwidth]{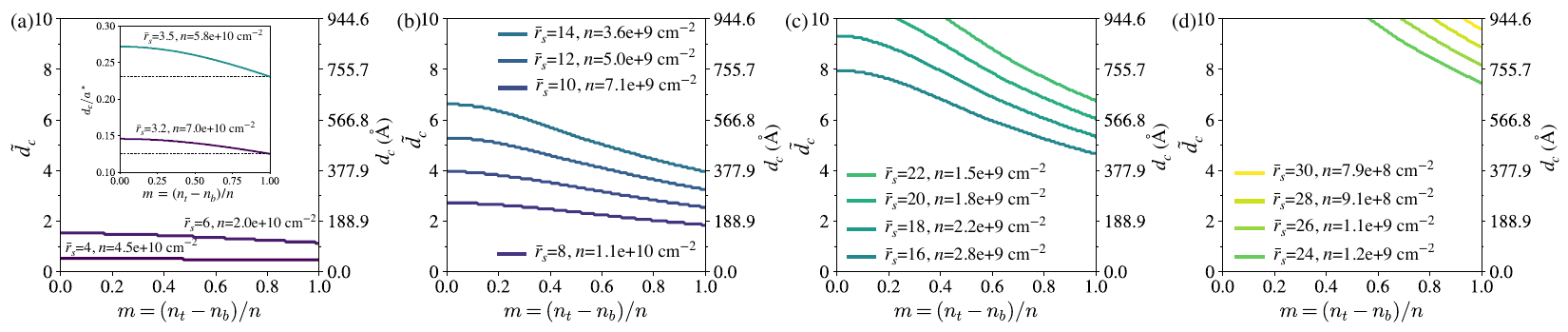}
\caption{\label{fig_dc_vs_m} {
Same as Fig.~\ref{fig_dc_vs_X}(a) but with more fixed values of $\bar{r}_s \in [4,30]$.
The inset in (a) shows two smaller $\bar{r}_s$.
At large density, corresponding to small $\bar{r}_s$,
$\tilde{d}_c$ exhibits a negligible dependence on $m$, as expected from our previous analysis in Eq.~(\ref{Eq_dc}). As $n \rightarrow n_{\rm max}$ in Eq.~(\ref{Eq_nmax}), $\tilde{d}_c$ is independent of $m$ and approaches zero. The black dashed lines in the inset of (a) plot the $\tilde{d}_c$ using Eq.~(\ref{Eq_dc}).
  }}
\end{figure*}

To clearly trace the evolution of $\tilde{d}_c$ with respect to tuning parameters $m$ and $\bar{r}_s$, we show in Fig.~\ref{fig_dc_vs_X}(a) $\tilde{d}_c$ versus $m$ for six values of $\bar{r}_s \in [3.8, 30]$, in Fig.~\ref{fig_dc_vs_X}(b) $\tilde{d}_c$ versus $\bar{r}_s$ for six values of $m \in [0,1]$. 
Figure~\ref{fig_dc_vs_m} complements the trend of $\tilde{d}_c$ with four additional plots that echo the configuration of Fig.~\ref{fig_dc_vs_X}(a), but with a refined set of $\bar{r}_s \in [4,30]$. 
In scenarios of high electron density, corresponding to low $\bar{r}_s$ as depicted in Fig.~\ref{fig_dc_vs_m}(a), we observe that $\tilde{d}_c$ exhibits a negligible dependence on $m$. This aligns with our previous analysis in Eq.~(\ref{Eq_dc}). The inset of Fig.~\ref{fig_dc_vs_m}(a) illustrates this trend that as $n \rightarrow n_{\rm max}$, $\tilde{d}_c$ is independent of $m$ and approaches zero.

\subsection{Partially polarized pseudospin state}
\label{subsec_PartialPolarized}
Previous subsections have been dedicated to examining interlayer coherent phases when the pseudospin is fully polarized, in the direction with polar angle $\theta=2\arctan(\alpha/\beta)$.
However, the assumption of complete pseudospin polarization may not always hold, distinct from the sharp paramagnetic to ferromagnetic phase transition in the Bloch transition. This subsection extends our investigation to the HF energies associated with states of partial pseudospin polarization.

\begin{figure*}[!htb]
\centering
\includegraphics[width=1.0\textwidth]{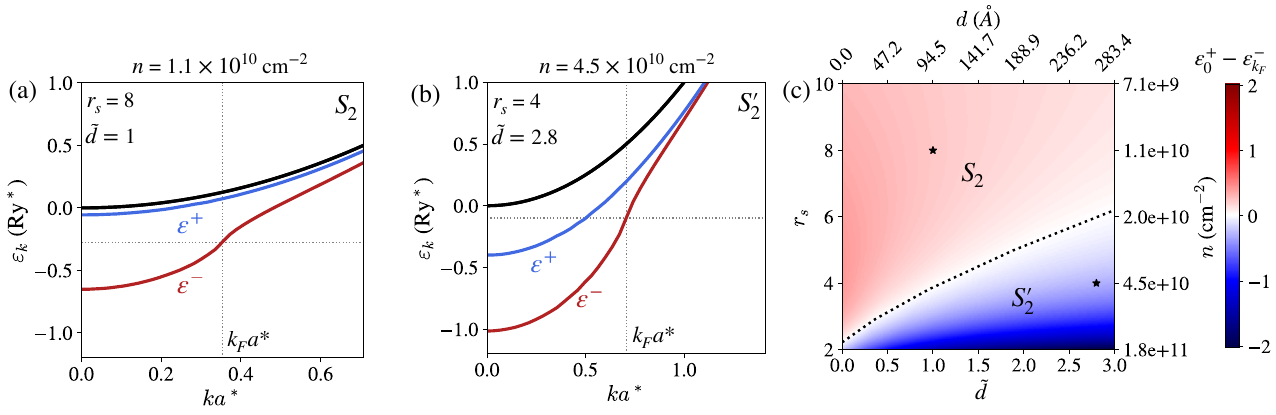}
\caption{\label{fig_QP_vals} { (a-b)
 Interlayer coherent quasiparticle energies in Eq.~(\ref{Eq_QP_energy}) for equal layer densities. $\varepsilon^{1,2}_\mathbf{k}$ are degenerate and represented by black solid lines. $\varepsilon^{-}_\mathbf{k}$ and $\varepsilon^{+}_\mathbf{k}$ are pseudospin polarized symmetric and anti-symmetric states, respectively. (a) $r_s=8$, $\tilde{d}=1$, corresponding to $n=1.1 \times 10^{10}$ cm$^{-2}$. (b) $r_s=4$, $\tilde{d}=2.8$, corresponding to $n=4.5 \times 10^{10}$ cm$^{-2}$. These two points are marked by stars in the phase diagram in (c). In (a-b), the Fermi momentum is denoted by the vertical dotted line and the corresponding Fermi energy is denoted by the horizontal one.
 We refer to case (a) as stable $S_2$ phase because only $|-\rangle$ band is occupied, and to (b) as the $S_2'$ phase in which both quasiparticle bands $|\pm \rangle$ are populated. (c) The phase diagram determined by the energy difference $\varepsilon^+_0 - \varepsilon^-_{k_F}$ as a function of $r_s$ and $\tilde{d}$.
 The $S_2$ phase in (c) encompasses a more extensive portion of the parameter space compared to the previously obtained phase diagram in Fig.~\ref{fig_phase_rsd}.
 The dashed line in (c) traces the phase transition boundary. 
  }}
\end{figure*}

The HF Hamiltonian of the spin polarized (majority spin $\sigma =\uparrow$) interlayer coherent state is
\begin{widetext}
\begin{equation}
\label{Eq_HMF}
\hat{H}^{S_\xi}_{\rm HF}(\mathbf{k}) = 
\begin{pmatrix}
\varepsilon_0(\mathbf{k})+V_{H,t}+V^{\uparrow \uparrow}_{x,t}(\mathbf{k}) & -\Delta_\mathbf{k}^{\uparrow \uparrow} & 0 & 0 \\
-(\Delta_\mathbf{k}^{\uparrow \uparrow})^* & \varepsilon_0(\mathbf{k})+V_{H,b}+V^{\uparrow \uparrow}_{x,b}(\mathbf{k}) & 0 & 0 \\
0 & 0 & \varepsilon_0(\mathbf{k})+V_{H,t} & 0 \\
0 & 0 & 0 & \varepsilon_0(\mathbf{k})+V_{H,b}
\end{pmatrix},
\end{equation}
\end{widetext}
its four eigenvalues are
\begin{equation}
\label{Eq_QP_energy}
\begin{split}
\varepsilon_\mathbf{k}^{1} &= \varepsilon_0(\mathbf{k}) + V_{H,t}, \\
\varepsilon_\mathbf{k}^{2} &= \varepsilon_0(\mathbf{k}) + V_{H,b}, \\
\varepsilon^{\pm}_{\mathbf{k}} &= \frac{1}{2}( \varepsilon_{t \mathbf{k}} + \varepsilon_{b \mathbf{k}} )
\pm \sqrt{ \xi^2_\mathbf{k} + |\Delta_\mathbf{k}|^2 },
\end{split}
\end{equation}
where 
\begin{equation}
\begin{split}
\varepsilon_{t \mathbf{k}} &= \varepsilon_0(\mathbf{k})+V_{H,t}+V^{\uparrow \uparrow}_{x,t}(\mathbf{k}), \\
\varepsilon_{b \mathbf{k}} &= \varepsilon_0(\mathbf{k})+V_{H,b}+V^{\uparrow \uparrow}_{x,b}(\mathbf{k}), \\
\xi_\mathbf{k} &= \frac{1}{2}( \varepsilon_{t \mathbf{k}} - \varepsilon_{b \mathbf{k}} ),
\end{split}
\end{equation}
and the HF potentials are in Eq.~(\ref{Eq_HFV_Sxi}).
The quasiparticle energy spectra are illustrated for two representative parameter sets: $r_s=8, \tilde{d}=1$ in Fig.~\ref{fig_QP_vals}(a) and $r_s=4, \tilde{d}=2.8$ in Fig.~\ref{fig_QP_vals}(b), corresponding to the two points marked by stars in the overall phase diagram in Fig.~\ref{fig_QP_vals}(c).
In Fig.~\ref{fig_QP_vals}(a-b), the Fermi momentum is denoted by the vertical dotted line and the corresponding Fermi energy is denoted by the horizontal one. Notably, in Fig.~\ref{fig_QP_vals}(b), both quasiparticle bands $\varepsilon^\pm$ are populated, which signals a deviation from the pseudospin fully polarized state, and we denote this phase as $S_2'$.

To crudely estimate the stability of this pseudospin partially polarized state, we compare the energy at zero momentum for the upper band, $\varepsilon^+_{0}$, with the energy at the Fermi momentum for the lower band $\varepsilon^-_{k_F}$, and show their difference across the ($r_s, \tilde{d}$) parameter space in Fig.~\ref{fig_QP_vals}(c). 
The red area corresponds to $\varepsilon^+_{0} > \varepsilon^-_{k_F}$, i.e., stable $S_2$ phase (as in Fig.~\ref{fig_QP_vals}(a)), and the blue area corresponds to $\varepsilon^+_{0} < \varepsilon^-_{k_F}$, which 
we refer to as $S_2'$ phase (the interlayer coherent phase characterized by two populated Fermi surfaces, as in Fig.~\ref{fig_QP_vals}(b)).
Compared to the previously obtained phase diagram in Fig.~\ref{fig_phase_rsd}, 
the $S_2'$ phase in Fig.~\ref{fig_QP_vals}(c) suggests that the interlayer coherent phase may persist over a broader range in the phase diagram than initially postulated when only fully polarized states were considered. But the most accurate phase diagram should be determined by self-consistent HF calculations which will be discussed in Sec.~\ref{sec_finiteT}.

\begin{figure*}[!htb]
\centering
\includegraphics[width=1.0\textwidth]{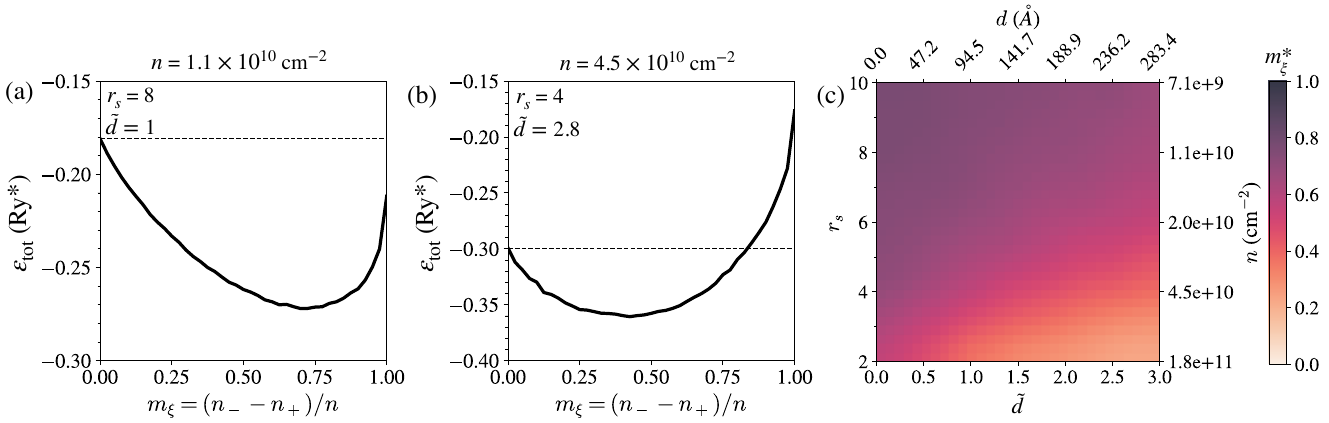}
\caption{\label{fig_E2FS_vs_m} {
(a-b) $\varepsilon_{\rm tot}^{S_2'}$ as a function of the pseudospin polarization $m_\xi$, at
 (a) $r_s=8$, $\tilde{d}=1$ and (b) $r_s=4$, $\tilde{d}=2.8$. The horizontal dotted lines mark the HF energies of the $S_1$ phase, converging to the same value of $\varepsilon_{\rm tot}^{S_2'}$ at $m_\xi=0$. In both (a) and (b),
 $\varepsilon_{\rm tot}^{S_2'}$ is minimized at an intermediate $m_\xi$. (c) The phase diagram of $m^*_\xi$ versus ($r_s$, $\tilde{d}$), where $m^*_\xi$ is the optimal polarization at which $\varepsilon^{S_2'}_{\rm tot}$ is minimized. $m^*_\xi$ tends to be $1$ for large $r_s$ and small $\tilde{d}$, while tends to be $0$ for small $r_s$ and large $\tilde{d}$.
  }}
\end{figure*}

To quantify the stability of this partially polarized state, we extend our exploration of the HF energy to the $S_2'$ phase, in which both quasiparticle states $|-,\mathbf{k} \rangle$ and  $|+,\mathbf{k} \rangle$ are occupied. 
The generalization of the HF energy of the $S_2$ phase in Eq.~(\ref{Eq_SP_energies}, \ref{Eq_SP_Etot1}, \ref{Eq_SP_Etot2}) to the two-Fermi-surface case is systematic and monotonic, as detailed below.

The kinetic, Hartree and exchange energies of the $S_2'$ state, expressed in both Fermi momenta and densities, are
\begin{widetext}
\begin{equation}
\begin{split}
E_{\rm kin} &= \frac{\hbar^2}{2m^*} \frac{A}{8\pi} \big( k_{F_-}^4 + k_{F_+}^4 \big) \\
&= \frac{\hbar^2}{2m^*} \frac{A}{8\pi} 16\pi^2 \big( n_-^2 + n_+^2 \big), \\
E_{\rm H}
&= \frac{e^2dA}{32 \epsilon \pi} (\alpha^2-\beta^2)^2 (k_{F_-}^2-k_{F_+}^2)^2 \\
&= \frac{e^2dA}{32 \epsilon \pi} (\alpha^2-\beta^2)^2 16\pi^2 (n_--n_+)^2, \\
E_{x}^{\rm intra} 
&= -\frac{e^2A}{\epsilon \pi^2} \Bigg[ \frac{1}{3}
(\alpha^4 + \beta^4)
(k^3_{F_-} +k^3_{F_+}) + \frac{1}{4} \alpha^2\beta^2 \bigg( k^3_{F_+} J(0,\frac{k_{F_-}}{k_{F_+}})  + k^3_{F_-} J(0,\frac{k_{F_+}}{k_{F_-}}) \bigg)
\Bigg], \\
E_{x}^{\rm inter} &= -\frac{e^2 A \alpha^2 \beta^2}{4\pi^2 \epsilon}
\Big[k_{F_-}^3 \big[J(k_{F_-}d,1) - J(k_{F_-}d,\frac{k_{F_+}}{k_{F_-}}) \big]
+ k_{F_+}^3 \big[J(k_{F_+}d,1) - J(k_{F_+}d,\frac{k_{F_-}}{k_{F_+}}) \big] \Big],
\end{split}
\end{equation}
\end{widetext}
where $k_{F_-}$ and $k_{F_+}$ are Fermi momenta, $n_-$ and $n_+$ are electron densities of quasiparticle bands $|-\rangle$ and $|+\rangle$ respectively.
The HF energy per electron is
\begin{equation}
\varepsilon_{\rm tot}^{S_2'} = \frac{1}{nA} \big( E_{\rm kin} + E_{\rm H} + E_{x}^{\rm intra} + E_{x}^{\rm inter} \big).
\end{equation}
Particularly, when $n_- = n_+ = n/2$, $k_{F_-}=k_{F_+}$, the pseudospin is unpolarized and the energies recover the ones of the $S_1$ state in Eq.~(\ref{Eq_EHF_S1}). 

In Fig.~\ref{fig_E2FS_vs_m}, we show $\varepsilon_{\rm tot}^{S_2'}$ as a function of pseudospin polarization $m_\xi$, defined as the ratio of density difference between $|- \rangle$ and $|+ \rangle$ eigenstates,
\begin{equation}
m_\xi = \frac{n_- - n_+}{n}.
\end{equation}
We specifically pick two points in ($r_s$, $\tilde{d}$) parameter space: one at $r_s=8, \tilde{d}=1$ in Fig.~\ref{fig_E2FS_vs_m}(a) where the ground state is anticipated to be $S_2$ ($m_\xi=1$), and the other at $r_s=8, \tilde{d}=2.8$ in Fig.~\ref{fig_E2FS_vs_m}(b) where $S_1$ state ($m_\xi=0$) is expected from the phase diagram Fig.~\ref{fig_phase_rsd}. In both Fig.~\ref{fig_E2FS_vs_m}(a) and Fig.~\ref{fig_E2FS_vs_m}(b), however, $\varepsilon^{S_2'}_{\rm tot}$ is minimized at an intermediate value of $m_\xi$. 
This optimal polarization, $m_\xi^*$, at which the energy is minimized, is depicted in Fig.~\ref{fig_E2FS_vs_m}(c). $m^*_\xi$ tends to be $1$ for large $r_s$ and small $\tilde{d}$, while tends to be $0$ for small $r_s$ and large $\tilde{d}$.

\begin{figure*}[!htb]
\centering
\includegraphics[width=0.9\textwidth]{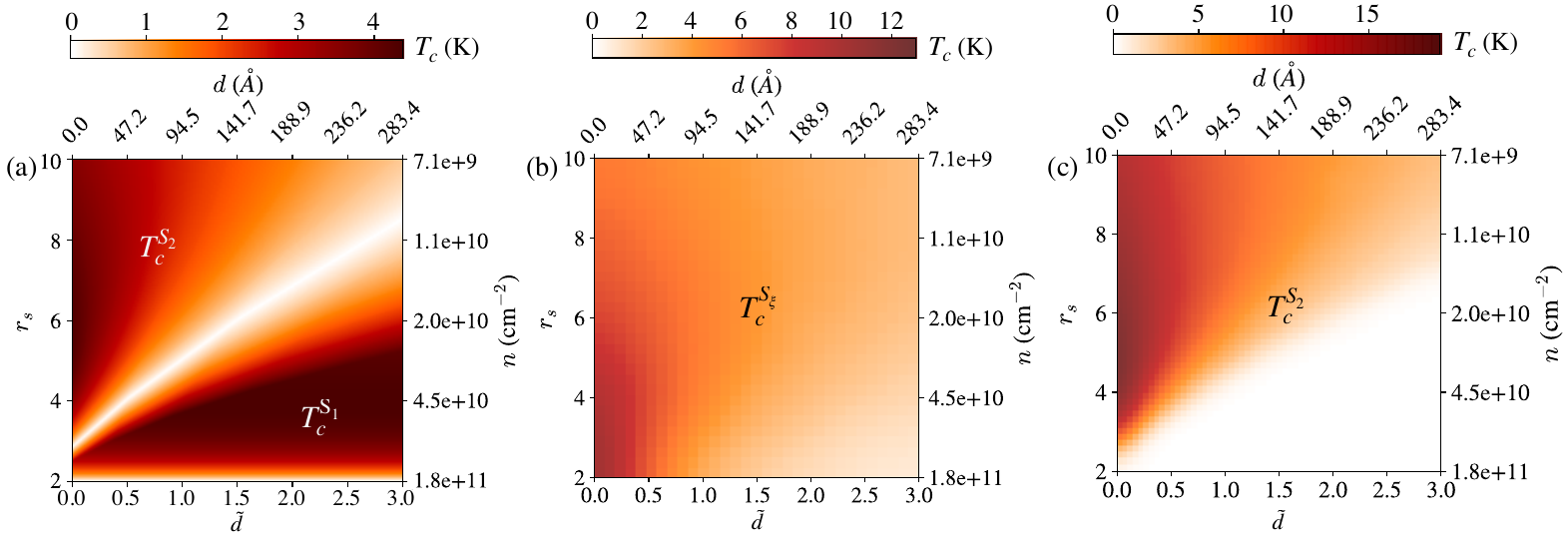}
\caption{\label{fig_Tc_phase} {
Critical temperature $T_c$ of the interlayer coherent phase. (a) $T_c$ is calculated by taking the energy difference between the ground state $\varepsilon_{\rm tot}^{G} = \min\{ \varepsilon^{S_0}_{\rm tot}, \varepsilon^{S_1}_{\rm tot}, \varepsilon^{S_2}_{\rm tot} \}$
and the second lowest energy state in Sec.~\ref{subsec_phase_equaln}.
(b) $T_c$ is calculated by taking the energy difference between the ground state $\varepsilon_{\rm tot}^{G} = \min\{ \varepsilon_{\rm tot}^{S_0}, \varepsilon_{\rm tot}^{S_1}, \varepsilon_{\rm tot}^{S_2}, \varepsilon_{\rm tot}^{S'_2}(m^*_\xi) \}$
and the second lowest energy state in Sec.~\ref{subsec_PartialPolarized}.
(c) $T_c$ is calculated by the finite-temperature self-consistent HF using the gap-like equation Eq.~(\ref{Eq_gap_ee}).
  }}
\end{figure*}

In Fig.~\ref{fig_Tc_phase}(b) we estimate the critical temperature $T_c$ by taking the difference between the ground state energy, $\varepsilon^{G}_{\rm tot}=$ min$\{ \varepsilon^{S_0}_{\rm tot}, \varepsilon^{S_1}_{\rm tot}, \varepsilon^{S_2}_{\rm tot}, \varepsilon^{S_2'}_{\rm tot}({m_\xi^*}) \}$, and the second lowest energy. Here $\varepsilon^{S_2'}_{\rm tot}({m_\xi^*})$ is the energy of the interlayer coherent state at the optimal pseudospin polarization $m^*_\xi$. For reference, $\varepsilon^{S_2'}_{\rm tot}(m_\xi = 1) = \varepsilon^{S_2}_{\rm tot}$ and  $\varepsilon^{S_2'}_{\rm tot}(m_\xi = 0) = \varepsilon^{S_1}_{\rm tot}$.
As a comparison, we show in Fig.~\ref{fig_Tc_phase}(a) the $T_c$ by taking the difference between the ground state energy, $\varepsilon^{G}_{\rm tot}=$ min$\{ \varepsilon^{S_0}_{\rm tot}, \varepsilon^{S_1}_{\rm tot}, \varepsilon^{S_2}_{\rm tot} \}$, and the second lowest energy. 
The $T_c$ phase diagram in Fig.~\ref{fig_Tc_phase}(b), incorporating pseudospin partially polarized states, yields a higher $T_c$—up to threefold—than that in Fig.~\ref{fig_Tc_phase}(a).

\section{Finite temperature phase diagrams}
\label{sec_finiteT}
In this section, we extend our study to the behavior of interlayer coherence at finite temperatures, by solving for the critical temperature $T_c$ using self-consistent HF approach.
We focus on the case for equal layer densities.
In Sec.~\ref{subsec_TcS2}, $T_c$ is determined for the interlayer coherent phase $S_2$.
In the subsequent Sec.~\ref{subsec_Tcexciton}, we broaden the scope of our examination to the exciton condensates in the e-h bilayer. By employing the same range of parameters as those in the e-e bilayer case, we facilitate a direct comparative analysis between the two systems' $T_c$ phase diagrams. This comparative framework not only highlights the unique characteristics of each system but also underscores the underlying different interacting nature governing their phase transitions at elevated temperatures.

\subsection{$T_c$ of interlayer coherent phase ($S_2$)}
\label{subsec_TcS2}
We focus on the majority spin subspace and ignore the minory spin for the spin polarized interlayer coherent phase $S_2$. To simplify notations, we will therefore ignore the spin superscripts in the following part. 
With spinor basis 
$(c_{t \mathbf{k}} \  c_{b \mathbf{k}})
^T$, the HF Hamiltonian is
\begin{equation}
\label{Eq_HS2}
\hat{H}^{S_2}_{\rm HF}(\mathbf{k}) = 
\begin{pmatrix}
\varepsilon_{t\mathbf{k}} & -\Delta_\mathbf{k} \\
-\Delta_\mathbf{k}^* & \varepsilon_{b\mathbf{k}}
\end{pmatrix},
\end{equation}
with eigenvectors
\begin{equation}
\begin{pmatrix}
+,\mathbf{k} \\
-,\mathbf{k}
\end{pmatrix}
= 
\begin{pmatrix}
\beta & -\alpha \\
\alpha & \beta
\end{pmatrix}
\begin{pmatrix}
c_{t \uparrow \mathbf{k}} \\
c_{b \uparrow \mathbf{k}} 
\end{pmatrix}
\end{equation}
and quasiparticle energies
\begin{equation}
\label{Eq_ee_QPenergy}
\begin{split}
\varepsilon^{\pm}_{\mathbf{k}} &= \frac{1}{2}( \varepsilon_{t \mathbf{k}} + \varepsilon_{b \mathbf{k}} )
\pm \sqrt{ \xi^2_\mathbf{k} + |\Delta_\mathbf{k}|^2 },
\end{split}
\end{equation}
where
\begin{equation}
\begin{split}
\varepsilon_{t \mathbf{k}} &= \varepsilon_0(\mathbf{k})+V_{H,t}+V_{x,t}(\mathbf{k}), \\
\varepsilon_{b \mathbf{k}} &= \varepsilon_0(\mathbf{k})+V_{H,b}+V_{x,b}(\mathbf{k}), \\
\xi_\mathbf{k} &= \frac{1}{2}( \varepsilon_{t \mathbf{k}} - \varepsilon_{b \mathbf{k}} ).
\end{split}
\end{equation}
At finite temperatures,
\begin{equation}
\label{Eq_SP_Vs}
\begin{split}
V_{H,t} &= \frac{2\pi e^2 d}{A \epsilon} \sum\limits_\mathbf{k} \big[\beta^2 f(\varepsilon^+_\mathbf{k}) + \alpha^2 f(\varepsilon^-_\mathbf{k}) \big], \\
V_{H,b} &= \frac{2\pi e^2 d}{A\epsilon}\sum\limits_\mathbf{k} \big[\alpha^2 f(\varepsilon^+_\mathbf{k}) + \beta^2 f(\varepsilon^-_\mathbf{k}) \big], \\
V_{x,t}(\mathbf{k}) &= -\frac{1}{A} \sum\limits_{\mathbf{k}'} V^{S}_{\mathbf{k}' - \mathbf{k}} \big[\beta^2 f(\varepsilon^+_{\mathbf{k}'}) + \alpha^2 f(\varepsilon^-_{\mathbf{k}'}) \big], \\
V_{x,b}(\mathbf{k}) &= -\frac{1}{A} \sum\limits_{\mathbf{k}'} V^{S}_{\mathbf{k}' - \mathbf{k}} \big[\alpha^2 f(\varepsilon^+_{\mathbf{k}'}) + \beta^2 f(\varepsilon^-_{\mathbf{k}'}) \big], \\
\Delta_k &
= \frac{1}{A} \sum\limits_{\mathbf{k}'} V^{D}_{\mathbf{k}'-\mathbf{k}} \alpha \beta \big[ f(\varepsilon^-_{\mathbf{k}'}) - f(\varepsilon^+_{\mathbf{k}'}) \big],
\end{split}
\end{equation}
where $f({\varepsilon_\mathbf{k}}) = [e^{(\varepsilon_{\mathbf{k}}-\mu)/k_{\rm B}T}+1]^{-1}$ is the Fermi-Dirac distribution and $\mu$ is the chemical potential determined by total density.
Both $\xi_k$ and $\Delta_k$, which are momentum-orientation independent, should be solved self-consistently by
\begin{equation}
\begin{split}
\xi_k &= \frac{\pi e^2}{A \epsilon} \sum\limits_{\mathbf{k}'} \Big(d - \frac{1}{|\mathbf{k}' - \mathbf{k}|} \Big) \\
& \qquad \qquad \quad \Big[(\alpha^2-\beta^2) \left(f(\varepsilon^-_{\mathbf{k}'})-f(\varepsilon^+_{\mathbf{k}'}) \right) \Big], \\
\Delta_k &= \frac{2\pi e^2}{A\epsilon} \sum\limits_{\mathbf{k}'} \frac{e^{-d|\mathbf{k}' - \mathbf{k}|}}{|\mathbf{k}' - \mathbf{k}|} \alpha \beta \big[ f(\varepsilon^-_{\mathbf{k}'}) - f(\varepsilon^+_{\mathbf{k}'}) \big].
\end{split}
\end{equation}
For the $S_2$ phase under our consideration, $\alpha = \beta = 1/\sqrt{2}$, and therefore $\xi_k = 0$. We just need to self-consistently solve the gap equation
\begin{equation}
\label{Eq_gap_ee}
\begin{split}
\Delta_k
&= \frac{\pi e^2}{A \epsilon} \sum\limits_{\mathbf{k}'} \frac{ e^{-d|\mathbf{k}-\mathbf{k}'|}}{|\mathbf{k}-\mathbf{k}'|} \big[ f(\varepsilon^-_{\mathbf{k}'}) - f(\varepsilon^+_{\mathbf{k}'}) \big].
\end{split}
\end{equation}

\begin{figure*}[!htb]
\centering
\includegraphics[width=0.7\textwidth]{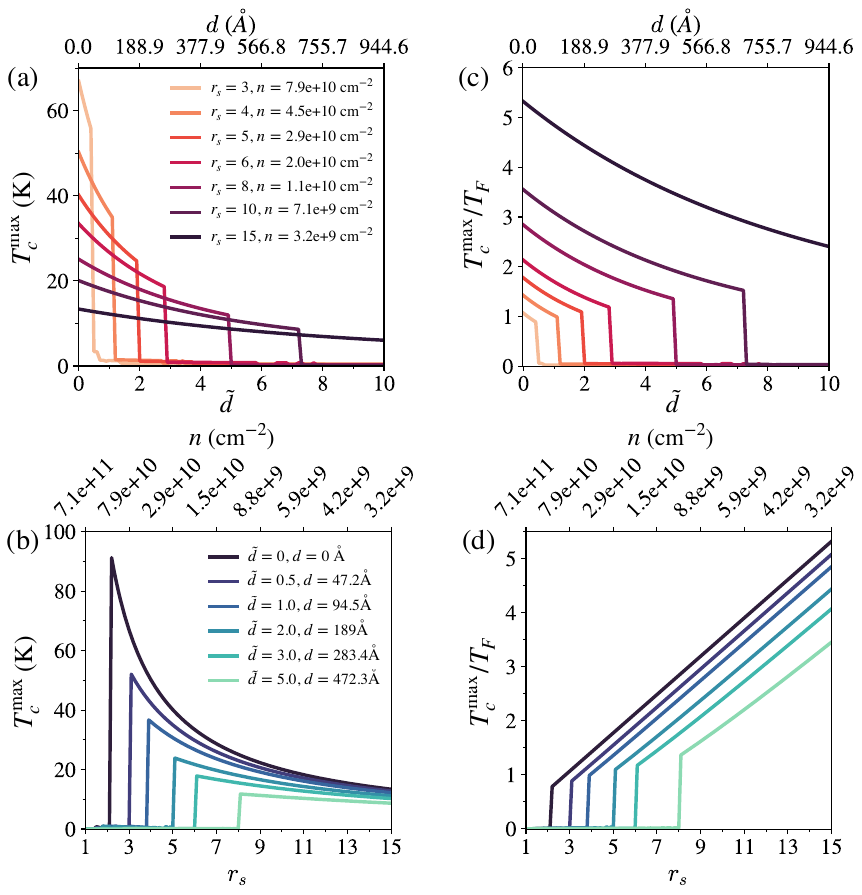}
\caption{\label{fig_ee_T0} {
Critical temperature of the $S_2$ phase in the e-e bilayer, obtained by zero-temperature self-consistent HF calculations.
$T_c^{\rm max}$ gives the upper-bound critical temperature, determined by $k_BT^{\rm max}_c =\max\limits_{k \leq k_c}\{\Delta_k\}$, where $k_c$ is the cutoff momentum and chosen to be $k_c = 2k_F$. 
(a) $T_c^{\rm max}$ as a function of $\tilde{d}$ for fixed $r_s$ values.
(b) $T_c^{\rm max}$ as a function of $r_s$ for fixed $\tilde{d}$ values.
(c) is the same as (a) but converted to the dimensionless $T_c^{\rm max}/T_F$, sharing the same legend as (a).
(d) is the same as (b) but converted to the dimensionless $T_c^{\rm max}/T_F$, sharing the same legend as (b).
 }}
\end{figure*}

At $T=0$, $f(\varepsilon^-_{\mathbf{k}}) - f(\varepsilon^+_{\mathbf{k}}) = 1$.
We first show the critical temperature obtained from zero-temperature self-consistent HF calculations, which gives the upper-bound $T^{\rm max}_c$ by $k_BT^{\rm max}_c =\max\limits_{k \leq k_c}\{\Delta_k\}$, where $k_c$ is the cutoff momentum and chosen to be $k_c = 2k_F$ in the following calculations.
We plot $T^{\rm max}_c$ versus $\tilde{d}$ in Fig.~\ref{fig_ee_T0}(a) and $T^{\rm max}_c$ versus $r_s$ in Fig.~\ref{fig_ee_T0}(b). There is always a critical $\tilde{d}_c$ ($r_{s,c}$) above (below) which the interlayer coherence vanishes. 
We convert $T_c^{\rm max}$ in Kelvin (obtained for GaAs e-e bilayers) to dimensionless $T^{\rm max}_c/T_F$ in Fig.~\ref{fig_ee_T0}(c-d), where $T_F$ characterizes the Fermi energy $k_BT_F = \varepsilon_F$.

\begin{figure*}[!htb]
\centering
\includegraphics[width=0.7\textwidth]{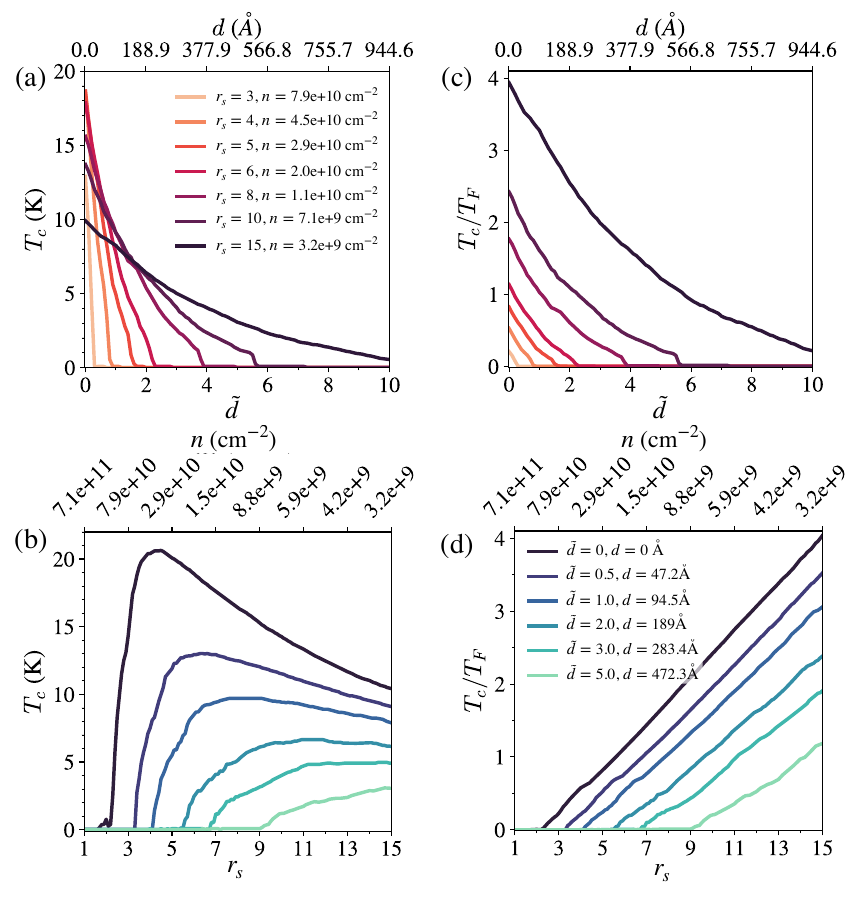}
\caption{\label{fig_ee_T} {
Critical temperature of the $S_2$ phase in the e-e bilayer, obtained by finite-temperature self-consistent HF calculations.
(a) $T_c$ as a function of $\tilde{d}$ for fixed $r_s$ values.
(b) $T_c$ as a function of $r_s$ for fixed $\tilde{d}$ values.
(c) is the same as (a) but converted to the dimensionless $T_c/T_F$, sharing the same legend as (a).
(d) is the same as (b) but converted to the dimensionless $T_c/T_F$, sharing the same legend as (b).
 }}
\end{figure*}

At finite temperatures, the critical temperature $T_c$ is determined by detecting the transition of $\Delta_k$ from $0$ to a finite value: for $T>T_c$, $\Delta_k=0$; for $T < T_c$, $\Delta_k > 0$.
The self-consistently calculated $T_c$ as a function of $\tilde{d}$ and $r_s$ are shown in Fig.~\ref{fig_ee_T}. 
As a function of $\tilde{d}$, $T_c$ at a large $r_s$ shows a power-law decay, qualitatively agreeing with the pseudospin stiffness behavior with respect to the interlayer separation in the quantum Hall bilayer at $\nu=1$.\cite{Moon_QHbilayer_1995}
Note that any mean-field calculation is crude for large layer separations, where the quantum fluctuations become important.
Same as the zero-temperature self-consistent HF calculations, there is a critical $\tilde{d}_c$ ($r_{s,c}$), above (below) which $T_c$ drops to zero, indicating the phase transition $S_2 \rightarrow S_1$.
As a function of $r_s$ shown in Fig.~\ref{fig_ee_T}(b), after entering the $S_2$ phase, $T_c$ initially rises quickly then drops slowly with $r_s$.
The behavior of the critical temperature becomes more apparent when expressed in terms of the dimensionless ratio $T_c/T_F$, which is depicted in Fig.~\ref{fig_ee_T}(c-d). Here $T_c/T_F$ decreases with $\tilde{d}$ following a power-law but increases almost linearly with $r_s$.

The complete critical temperature phase diagram, determined by the finite-temperature self-consistent HF calculations, is shown in Fig.~\ref{fig_Tc_phase}(c).
As expected, $T_c$ is maximized at small $\tilde{d}$ and moderate $r_s$ values. 
Notably, the phase boundary where $T_c$ vanishes closely mirrors that obtained using the zero-temperature HF energies in Fig.~\ref{fig_Tc_phase}(a). $T_c$ under finite-temperature self-consistent HF in Fig.~\ref{fig_Tc_phase}(c) is substaintially higher—by a factor of five—compared to the zero-temperature HF energy difference approach in Fig.~\ref{fig_Tc_phase}(a).
The magnitude of $T_c$ in Fig.~\ref{fig_Tc_phase}(c), however, aligns quantitatively with the estimation from zero-temperature HF energy difference approach when incorporating partially polarized pseudospins, in Fig.~\ref{fig_Tc_phase}(b).
Note that in Fig.~\ref{fig_Tc_phase}(c), $T_c$ vanishes for small $r_s$ and large $\tilde{d}$ because we restrict the self-consistent HF to the $S_2$ phase.


For comparative purposes, we show $T_c$ for the spin-polarized ferromagnetic phase $S_1$ calculated via finite-temperature self-consistent HF in Appendix~\ref{sec_appendixC} Fig.~\ref{fig_Tc_3D}(a).
The $S_1$ phase is the bilayer 2DEG with two equal Fermi surfaces with Fermi momentum $k_F = \sqrt{2\pi n}$. Therefore the $T_c$ of $S_1$ phase is $d$-independent. 
At $\tilde{d} = 0$, the $S_2$ phase (shown in the black lines in Fig.~\ref{fig_ee_T}(b,d)) is analogous to the $S_1$ phase, albeit with a single Fermi surface of momentum $k_F = \sqrt{4\pi n}$. 
Therefore, the trends observed for $T_c$ and $T_c/T_F$ at $\tilde{d}=0$ in Fig.~\ref{fig_ee_T}(b,d) share similar behaviors as those of the $S_1$ phase in  Fig.~\ref{fig_Tc_3D}(a), with differences attributable to the number of Fermi surfaces.

\begin{figure*}[!htb]
\centering
\includegraphics[width=0.7\textwidth]{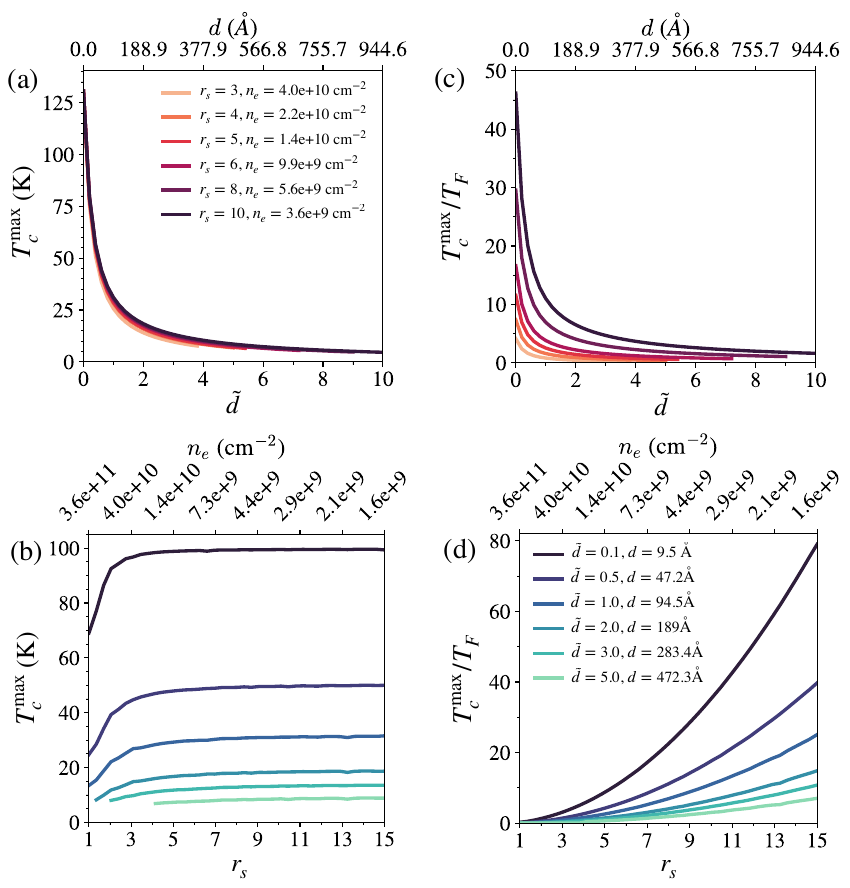}
\caption{\label{fig_exciton_T0} {
Critical temperature of exciton condensates in the e-h bilayer.
$T_c^{\rm max}$ is estimated by zero-temperature self-consistent HF calculations, which gives the upper-bound $T^{\rm max}_c$ by $k_BT^{\rm max}_c =\max\limits_{k \leq k_c}\{\Delta_k\}$, where $k_c$ is the cutoff momentum and chosen to be $k_c = 4$ nm$^{-1}$.
(a) $T_c^{\rm max}$ as a function of $\tilde{d}$ for fixed $r_s$ values.
(b) $T_c^{\rm max}$ as a function of $r_s$ for fixed $\tilde{d}$ values.
(c) is the same as (a) but converted to the dimensionless $T_c^{\rm max}/T_F$, sharing the same legend as (a).
(d) is the same as (b) but converted to the dimensionless $T_c^{\rm max}/T_F$, sharing the same legend as (b).
}}
\end{figure*}

\subsection{$T_c$ of exciton condensates in electron-hole bilayer}
\label{subsec_Tcexciton}
In this subsection we consider e-h bilayers, on an equal footing with the above-described e-e bilayers,  with electrons in one layer and holes in the other layer, forming exciton condensates. 
This architecture provides a direct comparison with the $S_2$ phase in e-e bilayers, facilitating a deeper and broader understanding of interlayer coherence phenomena.
For an intuitive picture, we categorize the electrons as residing in the conduction band (top layer) and the holes in the valence band (bottom layer).
The spinless HF Hamiltonian is
\begin{equation}
\begin{split}
\hat{\mathcal{H}}_{\rm HF} &= \sum\limits_{\mathbf{k}}
\begin{pmatrix}
c^\dagger_{c \mathbf{k}} & c^\dagger_{v \mathbf{k}}
\end{pmatrix} 
\begin{pmatrix}
\varepsilon_{c \mathbf{k}} & -\Delta_\mathbf{k} \\
-\Delta^*_\mathbf{k} & \varepsilon_{v \mathbf{k}}
\end{pmatrix}
\begin{pmatrix}
c_{c \mathbf{k}} \\
c_{v \mathbf{k}}
\end{pmatrix} \\
&= \sum\limits_{\mathbf{k}}
\begin{pmatrix}
\bar{\gamma}^\dagger_{\mathbf{k}} & \gamma^\dagger_{\mathbf{k}}
\end{pmatrix} 
\begin{pmatrix}
\varepsilon^{+}_{\mathbf{k}} & 0 \\
0 & \varepsilon^{-}_{\mathbf{k}}
\end{pmatrix}
\begin{pmatrix}
\bar{\gamma}_{\mathbf{k}} \\
\gamma_{\mathbf{k}}
\end{pmatrix}.
\end{split}
\end{equation}
The quasiparticle operators using the Bogoliubov transformation are
\begin{equation}
\begin{pmatrix}
+,\mathbf{k} \\
-,\mathbf{k}
\end{pmatrix}
\equiv
\begin{pmatrix}
\bar{\gamma}_\mathbf{k} \\
\gamma_\mathbf{k}
\end{pmatrix}
=
\begin{pmatrix}
u_\mathbf{k} & -v_\mathbf{k} \\
v^*_\mathbf{k} & u^*_\mathbf{k}
\end{pmatrix}
\begin{pmatrix}
c_{c \mathbf{k}} \\
c_{v \mathbf{k}}
\end{pmatrix},
\end{equation}
with quasiparticle energies
\begin{equation}
\begin{split}
\varepsilon^{\pm}_{\mathbf{k}} &= \frac{1}{2}( \varepsilon_{c \mathbf{k}} + \varepsilon_{v \mathbf{k}} )
\pm \sqrt{ \xi^2_\mathbf{k} + |\Delta_\mathbf{k}|^2 }, \\
\end{split}
\end{equation}
where 
\begin{equation}
\begin{split}
\varepsilon_{c \mathbf{k}} &= \varepsilon^{(0)}_{c \mathbf{k}} + V_{H,c} + V_{x,c}(\mathbf{k}), \\
\varepsilon_{v \mathbf{k}} &= \varepsilon^{(0)}_{v \mathbf{k}} + V_{H,v} + V_{x,v}(\mathbf{k}), \\
\varepsilon^{(0)}_{c \mathbf{k}} &= \frac{\hbar k^2}{2m^*}, \\
\varepsilon^{(0)}_{v \mathbf{k}} &= -\frac{\hbar k^2}{2m_c^*} -E_g,
\end{split}
\end{equation}
and
\begin{equation}
\label{Eq_epcv}
\begin{split}
\xi_\mathbf{k} &= \frac{1}{2}( \varepsilon_{c \mathbf{k}} - \varepsilon_{v \mathbf{k}} ), \\
\Delta_\mathbf{k} &= \frac{1}{A} \sum\limits_{\mathbf{k}'}V^D_{\mathbf{k}-\mathbf{k}'} \langle c^\dagger_{v\mathbf{k}'} c_{c\mathbf{k}'} \rangle. \\
\end{split}
\end{equation}
$E_g$ is the overlap between conduction and valence bands and is determined by the initial setting of the electron density $n_e$.
Note that we assign different effective masses to conduction and valence bands simply because the divergent negative Fermi sea should be taken into account in $m_v^*$.\cite{YPShim_excitonSOC_2009}
The valence band effective mass is renormalized by the exchange interactions of occupied remote band states:
\begin{equation}
\begin{split}
-\frac{\hbar^2 k^2}{2m_v^*} 
&= -\frac{\hbar^2 k^2}{2m^*} - \frac{1}{A} \sum\limits_{\mathbf{k}'} V^{S}_{\mathbf{k}-\mathbf{k}'} \langle c^\dagger_{v\mathbf{k}'} c_{v\mathbf{k}'} \rangle_0 \\
&= -\frac{\hbar^2 k^2}{2m^*} - \frac{1}{A} \sum\limits_{\mathbf{k}'} V^{S}_{\mathbf{k}-\mathbf{k}'} \rho_{vv}^0(\mathbf{k}')
\end{split}
\end{equation}
where $\rho_{vv}^0(\mathbf{k}') = \langle c^\dagger_{v\mathbf{k}'} c_{v\mathbf{k}'} \rangle_0 = 1$ is the expectation value in the reference state $|\Phi_0 \rangle$ that all valence band states are occupied and all conduction band states are empty. We should subtract this reference state expectation in all $\rho_{vv}(\mathbf{k})$ terms, and we denote them as
\begin{equation}
\tilde{\rho}_{vv}(\mathbf{k}) = \rho_{vv}(\mathbf{k}) - \rho^0_{vv}(\mathbf{k}).
\end{equation}

\begin{figure*}[!htb]
\centering
\includegraphics[width=0.7\textwidth]{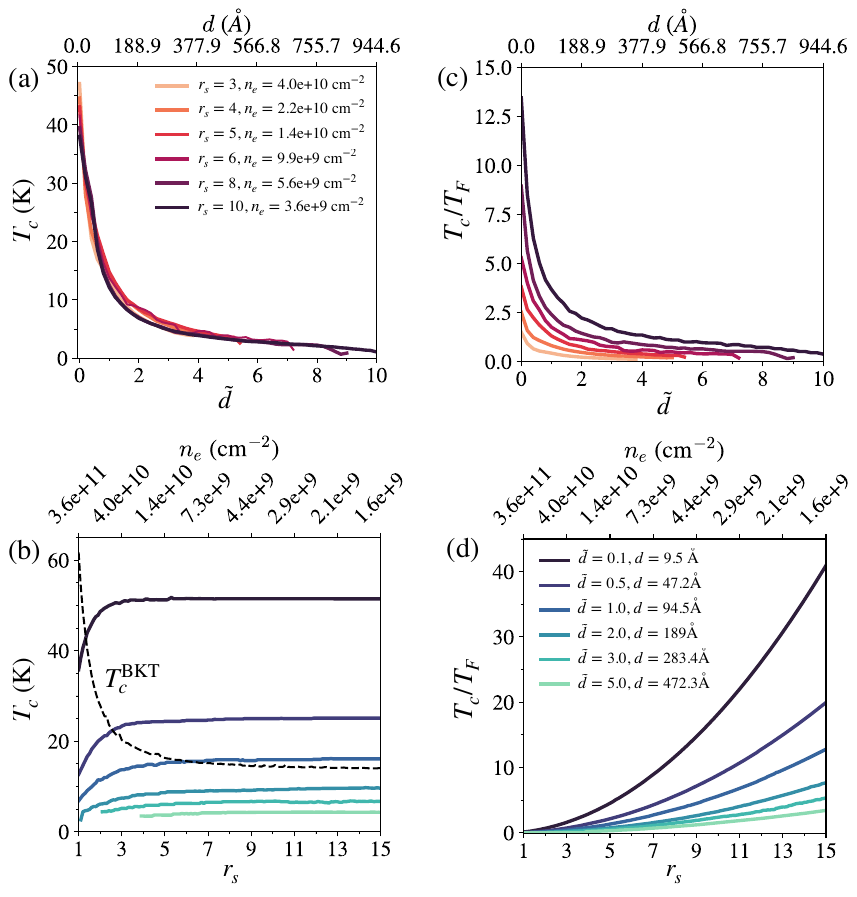}
\caption{\label{fig_exciton_T} {
Critical temperature $T_c$ of exciton condensates in the e-h bilayer.
$T_c$ is determined by the finite-temperature self-consistent HF calculations. 
(a) $T_c$ as a function of $\tilde{d}$ for fixed $r_s$ values.
(b) $T_c$ as a function of $r_s$ for fixed $\tilde{d}$ values.
(c) is the same as (a) but converted to the dimensionless $T_c/T_F$, sharing the same legend as (a).
(d) is the same as (b) but converted to the dimensionless $T_c/T_F$, sharing the same legend as (b).
The dashed line in (b) marks the critical temperature of the BKT transition, $T_c^{\rm BKT} \approx 1.3 \hbar^2n/2m^*$, at $\tilde{d}=0.1$, where $n$ is the converged exciton pair density.
  }}
\end{figure*}

The Hartree terms in Eq.~(\ref{Eq_epcv}) are
\begin{equation}
\begin{split}
V_{H,c} &= -V_{H,v} \\
&= \frac{\pi e^2 d}{A\epsilon} \sum\limits_{\mathbf{k}'} \Big( \rho_{cc}(\mathbf{k}') - \tilde{\rho}_{vv}(\mathbf{k}') \Big). \\
\end{split}
\end{equation}
For equal electron and hole densities that we consider here,
\begin{equation}
\begin{split}
V_{H,c} &= \frac{2\pi e^2 d n_e}{\epsilon}, 
\\
\end{split}
\end{equation}
where
\begin{equation}
n_e = \frac{1}{A} \sum\limits_{\mathbf{k}} \rho_{cc}(\mathbf{k})
\end{equation}
is the electron density.
The exchange terms in Eq.~(\ref{Eq_epcv}) are
\begin{gather}
V_{x,c} = -\frac{1}{A} \sum\limits_{\mathbf{k}'} V^S_{\mathbf{k}-\mathbf{k}'} \rho_{cc}(\mathbf{k}'), \nonumber\\
V_{x,v} = -\frac{1}{A} \sum\limits_{\mathbf{k}'} V^S_{\mathbf{k}-\mathbf{k}'} \tilde{\rho}_{vv}(\mathbf{k}'),\\
\Delta_\mathbf{k} = \frac{1}{A} \sum\limits_{\mathbf{k}'}V^D_{\mathbf{k}-\mathbf{k}'} \rho_{cv}(\mathbf{k}'). \nonumber
\end{gather}
At finite temperatures, the density matrix elements are
\begin{equation}
\begin{split}
\rho_{cc}(\mathbf{k}) &= \langle c^\dagger_{c\mathbf{k}} c_{c\mathbf{k}} \rangle = |v_\mathbf{k}|^2 f(\varepsilon^-_{\mathbf{k}}) + |u_\mathbf{k}|^2 f(\varepsilon^+_{\mathbf{k}}) \\
\tilde{\rho}_{vv}(\mathbf{k}) &= \langle c^\dagger_{v\mathbf{k}} c_{v\mathbf{k}} \rangle - 1 = |u_\mathbf{k}|^2 f(\varepsilon^-_{\mathbf{k}}) + |v_\mathbf{k}|^2 f(\varepsilon^+_{\mathbf{k}}) - 1 \\
\rho_{cv}(\mathbf{k}) &= \langle c^\dagger_{v\mathbf{k}} c_{c\mathbf{k}} \rangle = u^*_\mathbf{k} v_\mathbf{k} \left[ f(\varepsilon^-_{\mathbf{k}}) - f(\varepsilon^+_{\mathbf{k}}) \right].
\end{split}
\end{equation}
We need to self-consistently solve for
\begin{widetext}
\begin{equation}
\label{Eq_eh_sc}
\begin{split}
\xi_\mathbf{k} &= \frac{1}{2} \big( \varepsilon^{(0)}_{c\mathbf{k}} - \varepsilon^{(0)}_{v\mathbf{k}} \big) + \frac{2\pi e^2 d n_e}{\epsilon} - \frac{\pi e^2}{A\epsilon} \sum\limits_{\mathbf{k}'} \frac{1}{|\mathbf{k}-\mathbf{k}'|} \left[ 1-\frac{\xi_{\mathbf{k}'}}{\sqrt{\xi^2_{\mathbf{k}'}+|\Delta_{\mathbf{k}'}|^2}} 
\tanh \Big( \frac{\sqrt{\xi^2_{\mathbf{k}'}+|\Delta_{\mathbf{k}'}|^2}}{2k_BT} \Big)
\right], \\
\Delta_\mathbf{k}
&= \frac{\pi e^2}{A\epsilon} \sum\limits_{\mathbf{k}'} \frac{e^{-d|\mathbf{k}-\mathbf{k}'|}}{|\mathbf{k}-\mathbf{k}'|}
\frac{\Delta_{\mathbf{k}'}}{\sqrt{\xi^2_{\mathbf{k}'} + |\Delta_{\mathbf{k}'}|^2}} \tanh \Big( \frac{\sqrt{\xi^2_{\mathbf{k}'}+|\Delta_{\mathbf{k}'}|^2}}{2k_BT} \Big).
\end{split}
\end{equation}
\end{widetext}
We have used
\begin{align}
|u_\mathbf{k}|^2 &= \frac{1}{2} \left( 1+ \frac{\xi_\mathbf{k}}{\sqrt{\xi_\mathbf{k}^2 + |\Delta_\mathbf{k}|^2}} \right), \\
|v_\mathbf{k}|^2 &= \frac{1}{2} \left( 1- \frac{\xi_\mathbf{k}}{\sqrt{\xi_\mathbf{k}^2 + |\Delta_\mathbf{k}|^2}} \right), \\
u_\mathbf{k}^*v_\mathbf{k} &= \frac{\Delta_\mathbf{k}}{2\sqrt{\xi_\mathbf{k}^2 + |\Delta_\mathbf{k}|^2}}, \\
f(\varepsilon^-_{\mathbf{k}}) &- f(\varepsilon^+_{\mathbf{k}}) = \tanh \Big( \frac{\sqrt{\xi^2_\mathbf{k}+|\Delta_\mathbf{k}|^2}}{2k_BT} \Big).
\end{align}

At $T=0$, $f(\varepsilon^-_{\mathbf{k}}) - f(\varepsilon^+_{\mathbf{k}}) = 1$.
We first show the critical temperature obtained from zero-temperature self-consistent HF calculations, which gives the upper-bound $T^{\rm max}_c$ by $k_BT^{\rm max}_c =\max\limits_{k \leq k_c}\{\Delta_k\}$, where $k_c$ is the cutoff momentum and chosen to be $k_c = 4$ nm$^{-1}$ in the following calculations.
In Fig.~\ref{fig_exciton_T0}(a), we plot $T^{\rm max}_c$ versus $\tilde{d}$, observing an exponential decay of $T^{\rm max}_c$ with increasing $\tilde{d}$.
As a function of $r_s$ in Fig.~\ref{fig_exciton_T0}(b), $T^{\rm max}_c$ first rises and then almost stabilize at higher $r_s$ values. The approximate invariance in $T_c^{\rm max}$ at large $r_s$ is attributed to the converged exciton pair density and exciton binding energy.\cite{XZhu_exciton_1995}
We convert $T_c^{\rm max}$ in Kelvin (for GaAs bilayers) to dimensionless $T^{\rm max}_c/T_F$ in Fig.~\ref{fig_exciton_T0}(c-d).


Extending to finite temperatures, self-consistent HF calculations reveal $T_c$ trends similar to those at zero temperature. The $T_c$ obtained from finite-temperature calculations are approximately half of those predicted by $T_c^{\rm max}$, as depicted in Fig.~\ref{fig_exciton_T}, and is consistent with variational quantum Monte Carlo calculations.\cite{Neilson_ehMC_2018} Note that quantitative detailed differences between our mean-field study and the Monte Carlo study are attributed to the screening effects absent in our theory.
The dashed line in Fig.~\ref{fig_exciton_T}(b) marks the critical temperature of the BKT transition, $T_c^{\rm BKT} \approx 1.3 \hbar^2n/2m^*$, calculated using the converged exciton pair density $n$ at $\tilde{d}=0.1$. 
In the BEC limit (large $r_s$), $T_c^{\rm BKT}$ becomes independent of $r_s$ due to the converged exciton pair density.

\begin{figure*}[!htb]
\centering
\includegraphics[width=1.0\textwidth]{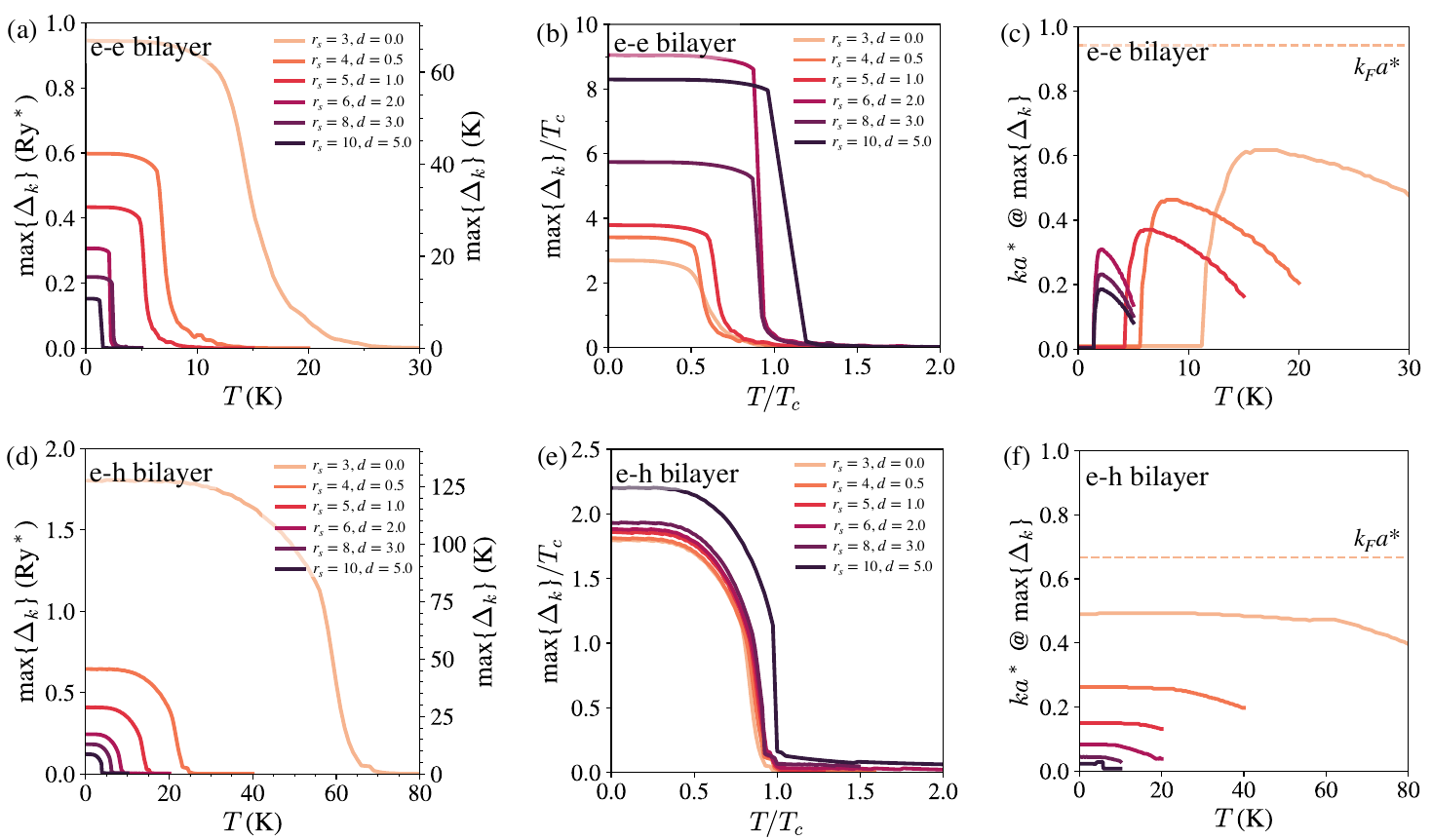}
\caption{\label{fig_Delta_vs_T} {
A side-by-side comparison of the interlayer coherence order parameter max$\{\Delta_k\}$ as a function of temperature in e-e (a-c) and e-h (d-f) bilayers.
(a,d) The order parameter max$\{\Delta_k\}$ versus temperature $T$. 
(b,e) Same as (a,d) but both max$\{\Delta_k\}$ and $T$ are converted to be dimensionless.
The fluctuations in the e-e bilayer is larger than in the e-h bilayer, because the e-e bilayer aligns more like the magnetic transition.
(c,f) The momentum $k$, at which the order parameter $\Delta_k$ is the maximum, versus the temperature.
For the e-e bilayer, maximum $\Delta_k$ is always at $k=0$ for $T<T_c$, while for the e-h bilayer, $\Delta_k$ is maximized at a finite momentum. This finite momentum decreases to zero as $r_s$ increases.\cite{XZhu_exciton_1995, XZhu_exciton_1996}
  }}
\end{figure*}

In Fig.~\ref{fig_Delta_vs_T}, we compare order parameters as a function of temperature for e-e and e-h bilayers. 
Panels (a) and (d) of Fig.~\ref{fig_Delta_vs_T}
display the maximum value of $\Delta_k$ against the temperature $T$.
For the e-e bilayer, we observe a relatively sharp drop in the order parameter as the temperature increases and approaches the critical value, followed by a more gradual decrease to zero with further temperature increase.
In contrast, the e-h bilayer exhibits a more gradual drop in the order parameter with increasing temperature, transitioning abruptly to zero as it approaches the critical temperature. 
This contrasting behavior between e-e and e-h bilayers is further elucidated in panels (b) and (e) of Fig.~\ref{fig_Delta_vs_T}, where both the order parameter and temperature are presented in dimensionless form.
Here, the fluctuations in the e-e bilayer are noticeably larger than those in the e-h bilayer, reflecting a closer resemblance to magnetic transition behaviors in the e-e bilayer.
Additionally, panels (c) and (f) of Fig.~\ref{fig_Delta_vs_T} trace the momentum $k$ at which the order parameter $\Delta_k$ reaches its maximum. In the e-e bilayer, $\Delta_k$ maximum consistently occurs at $k=0$ for temperatures below $T_c$, whereas in the e-h bilayer, $\Delta_k$  is maximized at a finite momentum. Notably, this finite momentum diminishes towards zero as $r_s$ increases.\cite{XZhu_exciton_1995, XZhu_exciton_1996}

\begin{figure*}[!htb]
\centering
\includegraphics[width=1.0\textwidth]{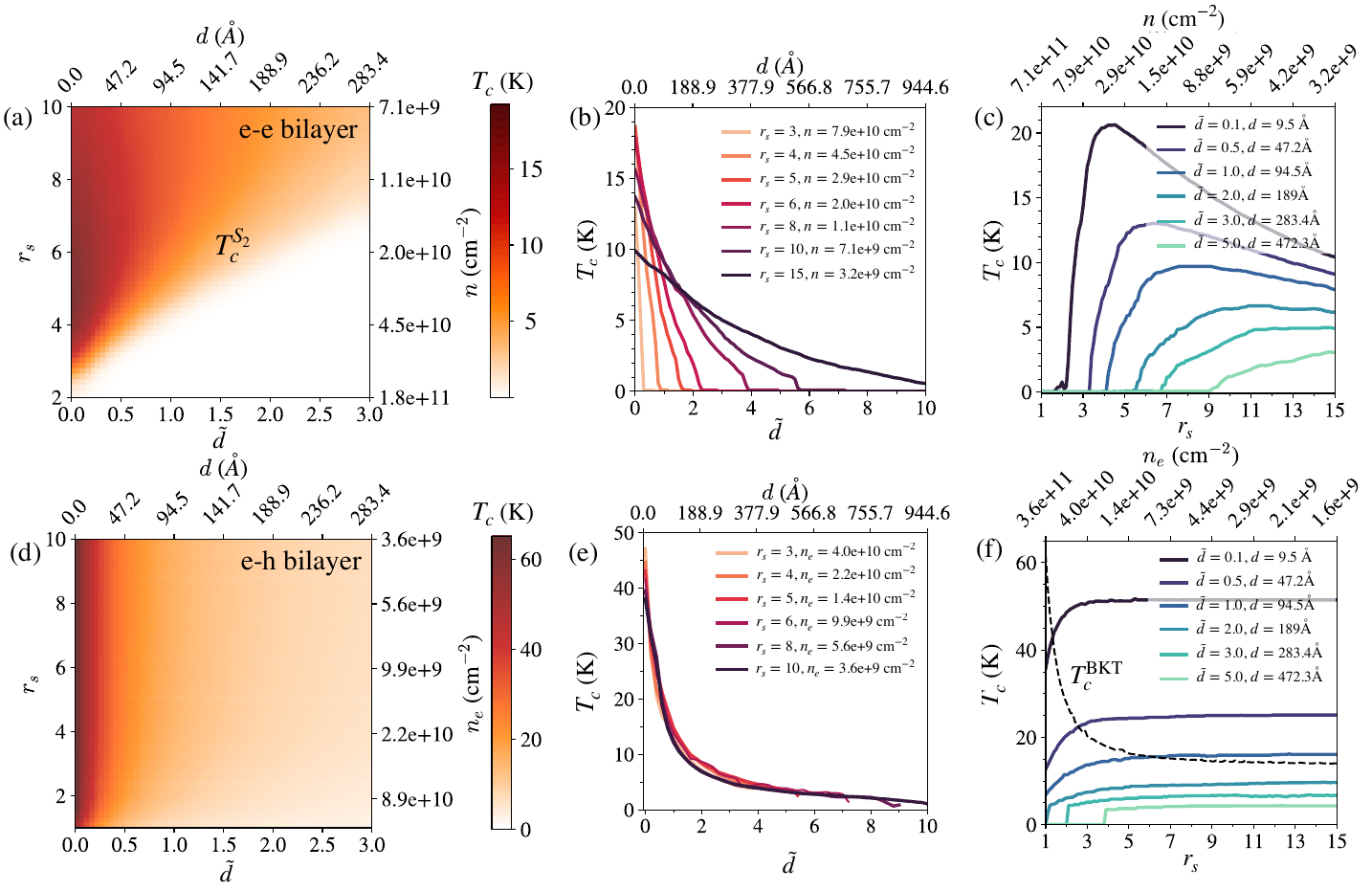}
\caption{\label{fig_Tc_compare} {
A side-by-side comparison of the $T_c$ behaviors in e-e (a-c) and e-h bilayer (d-f) systems.
The main difference between the e-e bilayer (XY pseudospin ferromagnet) and the e-h bilayer (exciton condensates) is that the former
necessitates a minimum $r_s$ value for its existence with the U(1) symmetry being broken only above a critical $r_{s,c}$, while the latter occurs for all $r_s$ at $T=0$. This is, however, not a practical difference because the e-h bilayer exciton condensates have exponentially low $T_c$ for small $r_s$, and therefore, it is unobservable for small $r_s$ any way.
  }}
\end{figure*}

We end our discussion with a side-by-side comparison of the $T_c$ behaviors in e-e and e-h bilayers in Fig.~\ref{fig_Tc_compare}. 
The main difference between XY pseudospin ferromagnetism in e-e bilayers and exciton condensates in e-h bilayers lies in their dependence of interlayer coherence on $r_s$. 
The e-e bilayer necessitates a minimum or critical $r_s$ for the existence of its coherent phase, with U(1) symmetry breaking only manifesting above a critical $r_{s,c}$. By contrast, the e-h bilayer exhibits exciton condensation for all $r_s$ values at zero temperature. 
However, this difference is not practically significant as exciton condensates in e-h bilayers exhibit exponentially low $T_c$ for small $r_s$ values, rendering them unobservable in this range.
Despite similarities in their gap-like equations (Eq.~(\ref{Eq_gap_ee}) and Eq.~(\ref{Eq_eh_sc})), e-e and e-h bilayers differ in two key aspects. 
First, e-h bilayers always include a Hartree term proportional to the layer separation $d$, a term that is absent in e-e bilayers with equal layer densities.
Second, at charge neutrality in e-h bilayers, where electron and hole densities are equal, the Fermi level invariably sits at the midpoint of the two quasiparticle bands. This results in the tangent term in the gap equation.
In contrast, in e-e bilayers, the term $(\varepsilon_{t\mathbf{k}}+\varepsilon_{b\mathbf{k}})$ in Eq.~(\ref{Eq_ee_QPenergy}) varies with $k$, meaning that  $f(\varepsilon^-_{\mathbf{k}'}) - f(\varepsilon^+_{\mathbf{k}'})$ in Eq.~(\ref{Eq_gap_ee}) cannot be simplified to the tangent function.
This comparison highlights the distinct thermal characteristics, underscoring the differences in the formation of coherent phases in these two analogous but distinct bilayer configurations.

\section{Effects of interlayer tunneling}
\label{sec_tunneling}
We focus in this section on the impact of interlayer tunneling on the XY pseudospin ferromagnetic transition in e-e bilayers, which is analogous to the influence of an in-plane magnetic field in ferromagnetic spin systems.
 
The HF Hamiltonian Eq.~(\ref{Eq_HS2}) is slightly modified with an interlayer tunneling $t$, acting as an effective magnetic field in the $x$-direction:
\begin{equation}
\hat{H}^{S_2}_{\rm HF}(\mathbf{k}) = 
\begin{pmatrix}
\varepsilon_{t\mathbf{k}} & -\Delta_\mathbf{k}+t \\
-\Delta_\mathbf{k}^*+t & \varepsilon_{b\mathbf{k}}
\end{pmatrix}.
\end{equation}
The quasiparticle energies become
\begin{equation}
\begin{split}
\varepsilon^{\pm}_{\mathbf{k}} &= \frac{1}{2}( \varepsilon_{t \mathbf{k}} + \varepsilon_{b \mathbf{k}} )
\pm \sqrt{ \xi^2_\mathbf{k} + |\Delta_\mathbf{k}-t|^2 },
\end{split}
\end{equation}
and the gap equation Eq.~(\ref{Eq_SP_Vs}) is modified in terms of Fermi-Dirac distribution.

\begin{figure*}[!htb]
\centering
\includegraphics[width=0.6\textwidth]{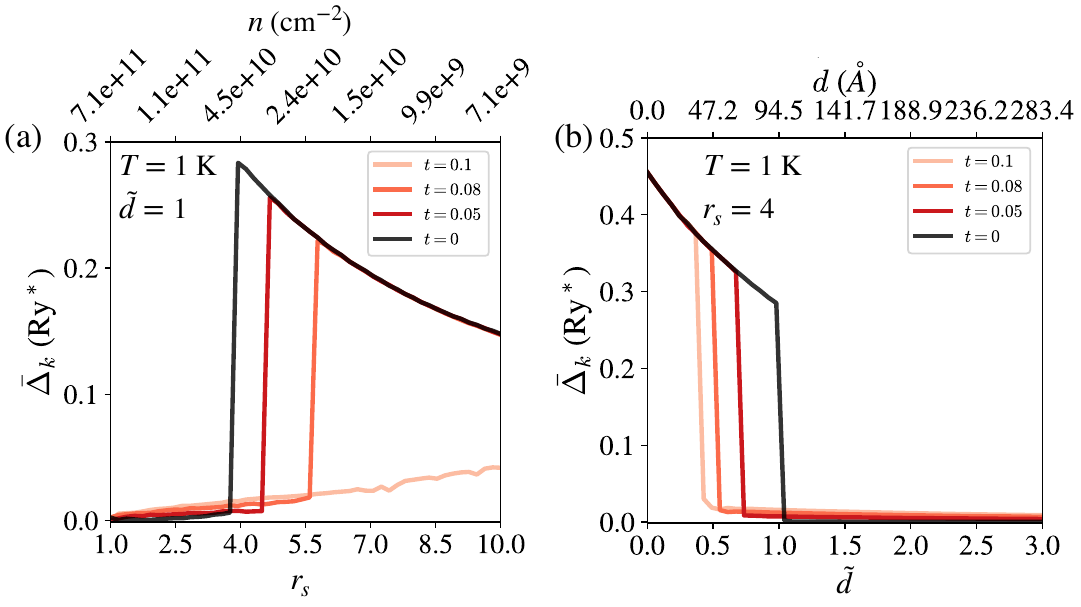}
\caption{\label{fig_tunneling} {
Effects of interlayer tunneling $t$ on the averaged spontaneous order parameter $\bar{\Delta}_k = \sum\limits_{k \in [0,k_c]} \Delta_{k} /N_k$, where $N_k$ is the number of 
$k$ points along $\mathbf{k} = k\hat{k}_x$ and $k \in [0,k_c]$, $k_c= 2k_F$ is the cutoff momentum.
Note that even $\bar{\Delta}_k=0$, two layers are still coherent, even though not spontaneously, because of the finite tunneling $t$. (a) $\bar{\Delta}_k$ versus $r_s$ at fixed $\tilde{d}=1$ and $T=1$ K. (b) $\bar{\Delta}_k$ versus $\tilde{d}$ at fixed $r_s=4$ and $T=1$ K.
The jumps in $\bar{\Delta}_k$ at critical $r_s$ and $\tilde{d}$ indicate the transition exactly as what happens for a ferromagnetic transition in a magnetic field.
In the presence of the interlayer tunneling $t$, the interlayer coherence transition becomes a crossover as the U(1) symmetry is explicitly broken by $t$, which in the pseudospin language is simply an applied magnetic field in the easy-plane defining the magnetization direction. 
  }}
\end{figure*}

We investigate the effects of a small interlayer tunneling $t$ on the averaged spontaneous gap $\bar{\Delta}_k = \sum\limits_{k \in [0,k_c]} \Delta_{k} /N_k$, where $N_k$ is the number of 
$k$ points along $\mathbf{k} = k\hat{k}_x$ and $k \in [0,k_c]$, $k_c= 2k_F$ is the cutoff momentum. Figure~\ref{fig_tunneling}(a) plots $\bar{\Delta}_k$ versus $r_s$ at fixed $\tilde{d}=1$ and temperature $T=1$ K, and Fig.~\ref{fig_tunneling}(b) plots $\bar{\Delta}_k$ versus $\tilde{d}$ at fixed $r_s=4$ and $T=1$ K.
The jumps in $\bar{\Delta}_k$ at critical $r_s$ and $\tilde{d}$ indicate the transition exactly as what happens for a ferromagnetic transition in a magnetic field.
As expected, the interlayer tunneling indeed supresses the spontaneous ferromagnetic ordering. This manifests as a shift in the critical $r_s$ to higher values and the critical $\tilde{d}$ to lower values, indicating a trend towards requiring a stronger coupling. 
In the presence of the interlayer tunneling $t$, the interlayer coherence transition becomes a crossover as the U(1) symmetry is explicitly broken by $t$, which in the pseudospin language is simply an applied magnetic field in the easy-plane defining the magnetization direction.

\section{Conclusion}
\label{sec_discussion}
In this paper, we carry out a comprehensive HF study of XY pseudospin ferromagnetism in zero-field e-e bilayers, and compare it with exciton condensation superfluid in zero-field e-h bilayers, for parabolic bands and long-range Coulomb interactions. 

At zero temperature, the phase diagram of e-e bilayers is determined by the HF energy.
In e-e bilayers with equal layer densities, the interlayer coherent phase with pseudospin ordered in the $xy$-plane is the stable ground state at lower electron densities (larger $r_s$) and reduced interlayer separation $d$.
There is, however, always a critical $r_s$ value for the interlayer coherence phase transition, and the system is a pseudospin paramagnet below this $d$-dependent critical $r_s$.
When layer densities are unequal, we find that the critical layer separation $d_c$, beyond which interlayer coherence vanishes, decreases with increasing layer density imbalance, but remains present even under complete layer polarization.
This layer polarization can be conceptualized as a pseudospin response to an effective magnetic field applied along the $z$-direction. Our results illustrate that, despite a strong effective magnetic field polarizing the pseudospin completely in the $z$-direction, the exchange-driven XY pseudospin ferromagnetic transition remains little affected.

At finite temperatures, we calculate the critical temperature $T_c$ using self-consistent HF theory. 
We find that $T_c$ for XY pseudospin ferromagnetism in e-e bilayers is approximately one-third of that for exciton condensates in e-h bilayers, indicating a comparatively weaker interlayer coherence in e-e bilayers. Fluctuations are larger in e-e bilayers because the phase transition aligns more closely to the magnetic transition.

Additionally, we evaluate the influence of weak interlayer tunneling on the interlayer coherence order parameter in e-e bilayers, mimicking the effects of an in-plane magnetic field on XY pseudospin ferromagnetism. In the presence of interlayer tunneling $t$, the interlayer coherence transition in e-e bilayers becomes a crossover, as the U(1) symmetry is explicitly broken by $t$.

A notable distinction between XY pseudospin ferromagnetism in e-e bilayers and exciton condensates in e-h bilayers is that the former requires a minimum $r_s$ value for its existence, with U(1) symmetry breaking only above a critical $r_{s}$ while the latter forms for any $r_s$ at zero temperature.
Practically, however, this difference is irrelevant because of the exponentially low $T_c$ for exciton condensates at small $r_s$ values, rendering them unobservable at high densities (i.e., the resultant $T_c$ is exponentially small although finite).
The fact that there is no critical density or equivalently no critical layer separation for the exciton condensation in bilayers was revealed in a quantum Hall bilayer experiment with the finding that strong interlayer e-h correlations exist even above the putative transition point showing the transition might be a fast BEC-BCS crossover and not a phase transition.\cite{Eisenstein_precursors_2019}
Both XY ferromagnetism and exciton condensates undergo finite-temperature BKT transitions and fall into the same universality class of phase transitions.
We emphasize that, unlike in quantum Hall bilayers with the Landau level filling being $1/2$ in each layer, the interlayer coherence phenomenon is physically very different in the zero-field e-e and e-h bilayers.  In the former case, the phenomenon is an XY pseudospin ferromagnetism problem whereas in the latter case, it is an interlayer excitonic condensation superfluid problem.  It just so happens that for quantum Hall bilayers, with the exact particle-hole symmetry for half-filled Landau levels, these two descriptions are precisely equivalent, but no such equivalence applies to the zero-field cases considered in the current work, where the exact particle-hole symmetry does not apply.
For quantum Hall bilayers, the existence or not of an interlayer separation tuned quantum phase transition between interlayer coherent (small separation)  and incoherent (large separation) phases has been much discussed in the lterature for more than 30 years using many different theoretical approaches.\cite{Fertig_QHBilayer_1989, Bonesteel_gauge_1996, Takao_compositefermion_1999, Kim_coulombdrag_2001, Moller_QHB_2009, Wagner_compositefermion_2021} We believe, based on our current work, that there is no zero temperature quantum phase transition in quantum Hall bilayers between interlayer coherent and incoherent phases as a function of layer separation (except at the trivial limit of infinite separation), and there is only a crossover between the strongly coupled BEC excitonic condensate at small separation to a weakly coupled BCS excitonic condensate at large separation.  Of course, the $T_c$ for the BCS phase being exponentially small and the system behaves like as if there is a phase transition around the crossover regime.

We mention that for very large $r_s$, each individual 2D layer would undergo a transition to a Wigner crystal phase (which is beyond the scope of our HF theory) breaking the translational invariance in each layer, but maintaining interlayer coherence, and such an interlayer-coherent intralayer Wigner crystal breaks both interlayer U(1) symmetry and intralayer translational symmetry, and is therefore a supersolid. Similar physics applies also for the e-h bilayers, where the large $r_s$ intralayer translational invariance breaking phase is an exciton crystal with the two layers being interlayer coherent breaking the interlayer U(1) symmetry, leading to an excitonic supersolid.\cite{SDS_WignerSupersolid_2006, DinhDuyVu_2023, JIALi_supersolid_2023}

Although this work lays the groundwork in understanding the differences in interlayer coherence between e-e and e-h bilayers by studying the simple 2DEG model, more complex 2D materials may exhibit richer phenomena. The interlayer coherence physics in these complex systems could be further altered by the band structure, topology, and screening effects.
The study of exciton condensates has been extensive in various bilayer systems, including recent graphene-based \cite{Tutuc_ehbilayer_2018} and transition-metal-dichalcogenide-based bilayers, \cite{Nguyen_ehbilayer_2023} as well as in double moir\'e superlattices. There is growing interest in exploring e-e pairing in these systems and its interplay with superconductivity. We will leave these investigations for future study.
We just mention in the passing that the well-established phenomena of canted antiferromagnetism in quantum Hall bilayer systems is an example of the interplay among interlayer coherence, spin physics, and interlayer tunneling in filled Landau levels.\cite{LZheng_SDS_1997, SDS_LZheng_1997, SDS_LZheng_1998, AHMacDonald_Jungwirth_1999, KunYang_1999, Eugene_SDS_1999, Schliemann_MacDonald_2000, Eugene_SDS_2000, Pellegrini_excitation_1997, Pellegrini_softmode_1998, ACGossard_2000}

Finally, we briefly discuss the experimental implications of interlayer coherence in zero-field e-h and e-e bilayers. It has been relatively clear that exciton BEC in quantum Hall bilayers is identifiable through specific transport measurements, such as quantized Hall drag resistance in the Coulomb drag geometry, vanishing longitudinal and Hall resistances in the counterflow geometry,\cite{Eisenstein_2DEGinB_2004, Tutuc_GaAs_exciton_2004, Nandi_CoulombDrag_2012, Eisenstein_CoulombDrag_2002, CDean_exciton_2022, Wiersma_activation_2004} and interlayer tunneling conductance anomalies.\cite{Spielman_QHBilayer_transport_2000, Spielman_QHBilayer_linearMode_2001, Eisenstein_Andreev_2011, Balents_QHB_tunneling_2001, Rossi_QHB_transport_2005}
Identifying exciton condensates in zero-field e-h bilayers remains challenging, though some experiments have come close.\cite{Seamons_CoulombDrag_2009,RuiRuiDu_exciton_2017,Tutuc_ehbilayer_2018, Davis_Josephson_ehbilayer_2023, Nguyen_ehbilayer_2023} 
Transport signatures for a diagnostic of exciton condensates in e-h bilayers include the Coulomb drag and counterflow resistances, and the in-plane Josephson effect.
For e-e bilayers, probing the XY pseudospin ferromagnetic metal might be feasible through transport measurements like interlayer tunneling in the presence of an in-plane magnetic field,\cite{Tutuc_hhbilayer_tunneling_2008, Eisenstein_eebilayer_tunneling_1991, Eisenstein_eebilayer_tunneling_1991_2}
and the pseudospin transfer torque.\cite{YKim_torque_2012}
In particular, we predict a giant interaction-induced  enhancement of the interlayer tunneling conductance peak in the interlayer coherent phase of e-e bilayers even if the non-interacting system has vanishing interlayer tunneling amplitude.  This giant interlayer tunnel resonance would be tunable by changing the layer density and will vanish above a critical density since the interlayer coherence in e-e bilayers can only happen above a critical $r_s$ value. 
In addition, another experimental approach to observe the spontaneous symmetry broken interlayer coherent phase could be studying the collective modes of the system, which should manifest the Goldstone mode in the symmetry broken phase, \cite{XYbilayer_Girvin_2000} and only regular plasmons in the normal bilayers. Thus, the collective modes of the system should differ for $T<T_c$ and $T > T_c$, with the former showing only one Goldstone mode, but the latter showing both  acoustic and optical plasmons of bilayers.\cite{SDS_CollectiveModes_1981, SDS_plasmon_1998} One issue here is that the Goldstone mode is likely to have a $q^{1/2}$ long wavelength dispersion (rather than being linear in momentum) because of the 2D analog of the Anderson-Higgs mechanism.\cite{SDS_collectivemodes_1990}

Our specific predictions in this work on the dependence of the interlayer coherent phase on the layer density imbalance and/or weak interlayer tunneling should be a useful guide for experimental explorations of the predicted bilayer coherence.  
The fact that the corresponding transition has been extensively experimentally studied in quantum Hall bilayers, where the XY pseudospin ferromagnet and the exciton superfluid become the same, gives considerable hope that both XY pseudospin ferromagnetism and exciton superfluid condensation will soon be experimentally observed at zero magnetic field in high quality bilayer systems.

Lastly, we comment on the limitations of the HF mean-field theory.
Because of its lack of dynamical screening effects and fluctuations,
the HF theory is not a quantitatively accurate method to predict the symmetry broken ground state, and it overestimates $T_c$ of the phase transition at finite temperatures. However, the HF approximation should always provide a qualitatively accurate phase diagram since it is a mean-field theory using realistic (i.e., Coulomb in our case) interactions. The HF mean-field theory is used extensively in the literature to obtain the quantum phase diagram of electronic materials and is always the first theory to provide a guide to experiments.
In addition, our HF theory in predicting the pseudospin phase transitions is proved to be accurate on the mean-field level, as demonstrated in a parallel study using a mean-field self-consistent iterative approach.\cite{Tessa_2024}
In the single-layer 2DEG, the spontaneous spin polarization actually occurs at a much larger critical $r_s$ value than the Bloch transition $r_{s,B} \sim 2$ if correlations, which are absent in the HF theory, are included.
Given that the pseudospin polarized (or pseudospin ferromagnetic) phase predicted in our work is a direct result of the interlayer exchange interaction, the pseudospin polarized phase will also be suppressed by correlations. Therefore, we should expect all the pseudospin phase transtitions predicted in our phase diagrams occur at larger $r_s$ and smaller $d$.
But we see no reason for our calculated phase diagram to be qualitatively incorrect although the symmetry broken phases are often overestimated in the HF theory.  

Finally, we mention that our work is directly applicable to 2D GaAs bilayers (as emphasized by all our figures showing results for GaAs system parameters) where the 2D Fermi surface is not multiply-connected, and  is in general isotropic and parabolic.  For more complicated 2D Fermi surfaces (e.g. transition metal dichalcogenide layers, moir\'e materials, multilayer graphene systems), our work would remain qualitatively, but not quantitatively valid, since symmetry breaking transitions are driven by exchange interactions which always dominate the kinetic energy at low enough densities in Coulomb coupled systems independent of the Fermi surface details.

\section{Acknowledgments}
This work is supported by the Laboratory for Physical Sciences.

\appendix
\section{Summary of useful summations and integrals}
\label{Appendix_integrals}
We summarize some useful summations and integrals which are frequently used in the equations in the main text.
\begin{align}
\sum\limits_{k \leq k_F} &= \frac{A}{4\pi} k_F^2, \\
\sum\limits_{k \leq k_F} k^2 &= \frac{A}{8\pi} k_F^4, 
\end{align}
\begin{widetext}
\begin{align}
\sum\limits_{k,k' \leq k_F} \frac{1}{|\mathbf{k} - \mathbf{k}'|} &= \frac{A}{(2\pi)^2} \sum\limits_{k \leq k_F} \int_{k' \leq k_F} d^2\mathbf{k}' \frac{1}{|\mathbf{k} - \mathbf{k}'|} \nonumber\\
&= \frac{A}{(2\pi)^2} \sum\limits_{k \leq k_F} \int_0^{k_F} dk' k' \int_0^{2\pi} \frac{d\theta}{\sqrt{(k')^2 + k^2 - 2kk' \cos\theta}} \nonumber\\
&= \frac{A}{(2\pi)^2} \sum\limits_{k \leq k_F} 4 k_F E\Big(\frac{k}{k_F} \Big) \nonumber\\
&= \frac{A^2 k_F^3}{2\pi^3} \int_0^1 dx xE(x) \nonumber\\
&= \frac{A^2 k_F^3}{3\pi^3}, 
\end{align}
\end{widetext}
\begin{widetext}
\begin{align}
\sum\limits_{k' \leq k_F} \frac{1}{|\mathbf{k} - \mathbf{k}'|} &= \frac{A}{(2\pi)^2} \int_0^{k_F} dk' k' \int_0^{2\pi} \frac{d\theta}{\sqrt{(k')^2 + k^2 - 2kk' \cos\theta}} \nonumber\\
&= \frac{A}{\pi^2} k_F 
f_{2D} \left(\frac{k}{k_F} \right), \\
\sum\limits_{k,k' \leq k_F} \frac{e^{-|\mathbf{k} - \mathbf{k}'|d}}{|\mathbf{k}-\mathbf{k}'|} &= \frac{A^2}{(2\pi)^3} \int_0^{k_F} dk k  \int_0^{k_F} dk' k' \int_0^{2\pi} d\theta \frac{e^{-d \sqrt{(k')^2+k^2-2k'k\cos \theta}}}{\sqrt{(k')^2+k^2-2k'k\cos \theta}} \nonumber\\
&= \frac{A^2k_F^3}{(2\pi)^3}  \int_0^1 dx x \int_0^1 dy y \int_0^{2\pi} d\theta \frac{e^{-k_Fd \sqrt{x^2+y^2-2xy\cos \theta}}}{\sqrt{x^2+y^2-2xy\cos \theta}} \nonumber\\
&=\frac{A^2k_F^3}{(2\pi)^3} J \left(k_Fd \right),\\
\sum\limits_{k' \leq k_F} \frac{e^{-|\mathbf{k} - \mathbf{k}'|d}}{|\mathbf{k}-\mathbf{k}'|} &= \frac{A}{(2\pi)^2}  \int_0^{k_F} dk' k' \int_0^{2\pi} d\theta \frac{e^{-d \sqrt{(k')^2+k^2-2k'k\cos \theta}}}{\sqrt{(k')^2+k^2-2k'k\cos \theta}} \nonumber\\
&= \frac{Ak_F}{(2\pi)^2} \int_0^1 dy y \int_0^{2\pi} d\theta \frac{e^{-k_Fd \sqrt{x^2+y^2-2xy\cos \theta}}}{\sqrt{x^2+y^2-2xy\cos \theta}} \nonumber\\
&=\frac{Ak_F}{(2\pi)^2} I \left(x, k_Fd \right),
\end{align}
\end{widetext}
where 
\begin{equation}
f_{2D}(x) = 
\begin{cases} 
E(x), & x \leq 1, \\
x \left[ E\left(\frac{1}{x}\right) - \left(1 - \frac{1}{x^2}\right) K\left(\frac{1}{x}\right) \right], & x \geq 1,
\end{cases}
\end{equation}
$K(x)$ and $E(x)$ are the complete elliptic integral of the first and the second kind respectively. $E(x)$ can be expressed as a power series
\begin{equation}
\begin{split}
E(x) = \frac{\pi}{2} \sum\limits_{n=0}^\infty \Bigg( \frac{(2n)!}{2^{2n} (n!)^2} \Bigg)^2 \frac{x^{2n}}{1-2n},
\end{split}
\end{equation}
and the integration 
\begin{equation}
\begin{split}
\int_0^1 dx xE(x) &= \frac{\pi}{2} \sum\limits_{n=0}^\infty \Bigg( \frac{(2n)!}{2^{2n} (n!)^2} \Bigg)^2 \frac{1}{2(n+1)(1-2n)} \\
&= \frac{2}{3}.
\end{split}
\end{equation}
The triple integral $J(k_Fd)$
\begin{equation}
\begin{split}
J(k_Fd) = \int_0^1 dx x \int_0^1 dy y &\int_0^{2\pi} d\theta \\
&\frac{e^{-k_Fd \sqrt{x^2+y^2-2xy \cos \theta}}}{\sqrt{x^2+y^2-2xy \cos \theta}}
\end{split}
\end{equation}
and the integral
\begin{equation}
I(x,k_Fd) = \int_0^1 dy y \int_0^{2\pi} d\theta \frac{e^{-k_Fd \sqrt{x^2+y^2-2xy\cos \theta}}}{\sqrt{x^2 + y^2-2xy\cos \theta}}
\end{equation}
should be evaluated numerically.

\section{HF energies of the four competing states}
\label{HF_energy_T0}
\begin{figure*}[!htb]
\centering
\includegraphics[width=1.0\textwidth]{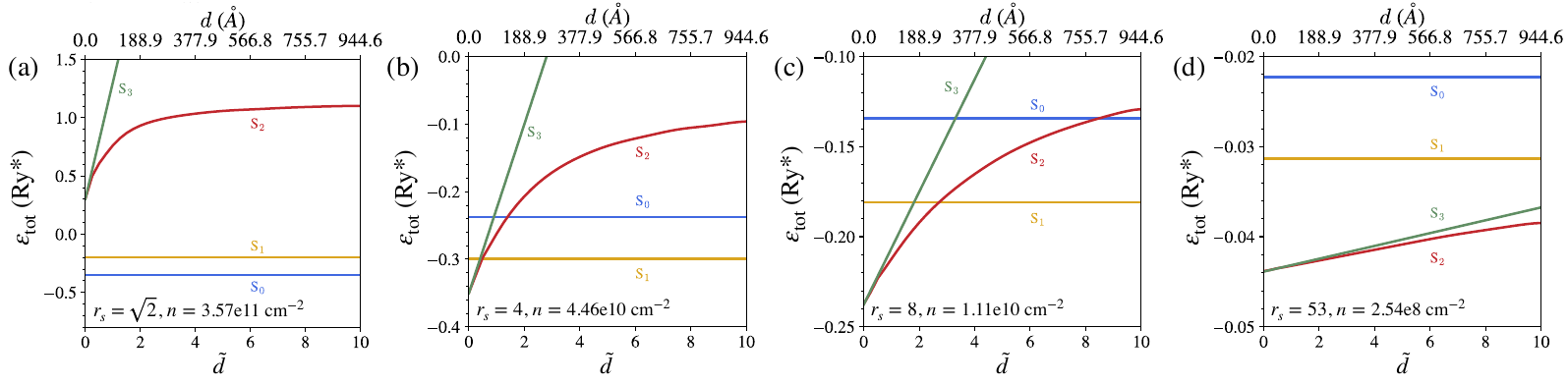}
\caption{\label{fig_Evsd_T0} {
 HF energies as a function of layer distance $\tilde{d}$ for (a) $r_s=\sqrt{2}$, (b) $r_s = 4$, (c) $r_s = 8$ and (d) $r_s=53$. Dimensionless $r_s$ and $\tilde{d}$ are also converted to density $n$ in unit of cm$^{-2}$ and $d$ in unit of $\text{\AA}$, using GaAs quantum well parameters.
  }}
\end{figure*}

In Fig.~\ref{fig_Evsd_T0}, we show HF energies for competing ground states $S_0$, $S_1$, $S_2$ and $S_3$ as a function of layer distance $\tilde{d}$ for specific $r_s$ values. The $S_2$ phase always has a lower energy than the $S_3$ phase except at $\tilde{d}=0$.

\begin{figure*}[!htb]
\centering
\includegraphics[width=0.9\textwidth]{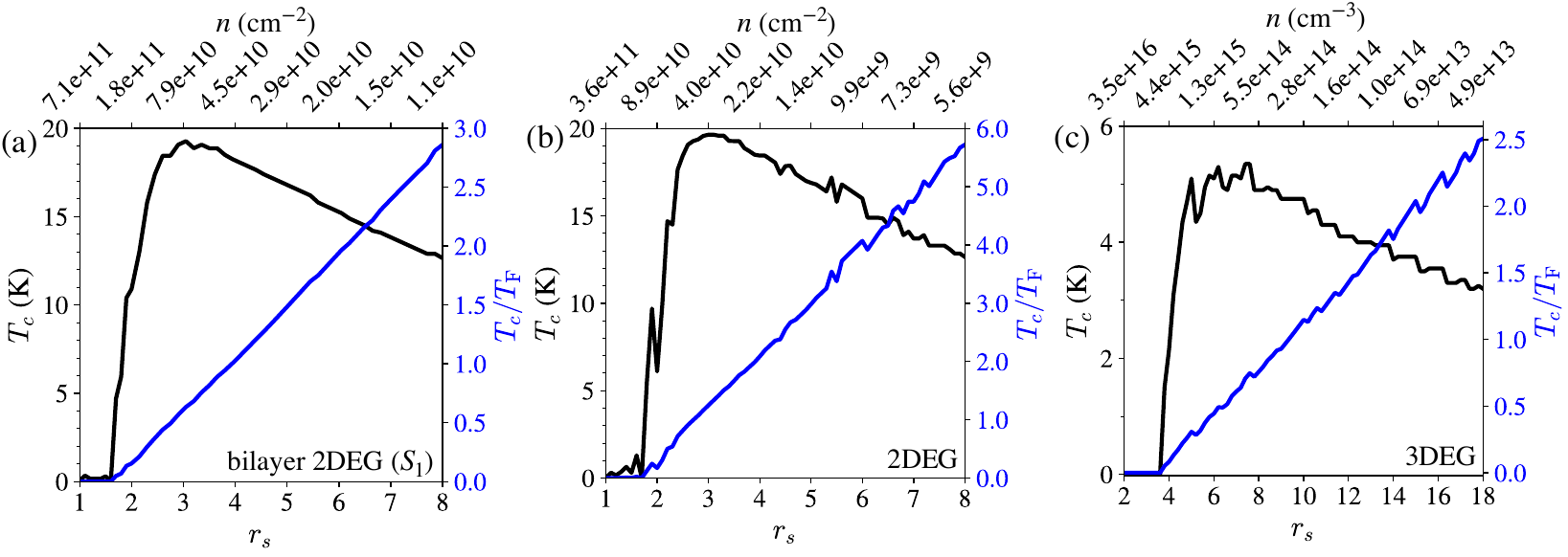}
\caption{\label{fig_Tc_3D} {
$T_c$ of the spin polarized state, calculated by finite-temperature self-consistent HF. (a) For bilayer 2DEG, i.e., the $S_1$ phase in Sec.~\ref{subsec_S1}. Note that $T_c$ of the $S_1$ phase is $d$-independent. (b) For single 2DEG. (c) For single 3DEG.
  }}
\end{figure*}

\section{$T_c$ of the spin polarized but pseudospin unpolarized phase ($S_1$)}
\label{sec_appendixC}

To compare with the $T_c$ of bilayer 2DEG $S_1$ phase in Sec.~\ref{subsec_S1} and in Fig.~\ref{fig_ee_T}(e,f), we show $T_c$ of the spin polarized phase in single 2DEG and single 3DEG.

\subsection{2DEG}
For the spin polarized phase, the eigenenergy of the majority spin is
\begin{equation}
\varepsilon_{\mathbf{k}} = \varepsilon_0(\mathbf{k}) + V_{x}(\mathbf{k}).
\end{equation}
where
\begin{equation}
\begin{split}
\varepsilon_0(\mathbf{k}) &= \frac{\hbar^2 k^2}{2m} \\
V_{x}(\mathbf{k}) &= -\frac{1}{A} \sum\limits_{\mathbf{k}'} \frac{2\pi e^2}{\epsilon |\mathbf{k}-\mathbf{k}'|} f(\varepsilon_{\mathbf{k}'})
\end{split}
\end{equation}
The occupied minority spin state has energy $\varepsilon_0(\mathbf{k})$. 
At $T=0$, the Fermi-Dirac distribution $f(\varepsilon_{\mathbf{k}})$ is the step function, and
\begin{equation}
V_{x}(\mathbf{k}) = -\frac{4a^*}{\pi}k_F E(\frac{k}{k_F})
\end{equation}
where Fermi momentum $k_F = \sqrt{2\pi n}$.
$T_c$ of the 2DEG spin polarized state, calculated by finite-temperature SCHF, is shown in Fig.~\ref{fig_Tc_3D}(b).

\subsection{3DEG}
For the 3DEG, the eigenenergy of the majority spin is 
\begin{equation}
\begin{split}
\varepsilon_{\mathbf{k}} &= \varepsilon_0(\mathbf{k}) + V_{x}(\mathbf{k}),
\end{split}
\end{equation}
where
\begin{equation}
\begin{split}
\varepsilon_0(\mathbf{k}) &= \frac{\hbar^2 k^2}{2m} ,\\
V_{x}(\mathbf{k}) &= - \frac{1}{L^3} \sum\limits_{\mathbf{k}'} \frac{4\pi e^2}{\epsilon |\mathbf{k} - \mathbf{k}'|^2} f(\varepsilon_{\mathbf{k}'}),
\end{split}
\end{equation}
and the Fermi momentum $k_F = (6\pi^2 n)^{1/3}$.
The dimensionless distance
\begin{equation}
r_s a^* = \left(\frac{3}{4\pi n} \right)^{1/3}.
\end{equation}
At $T=0$, the exchange potential 
\begin{equation}
\begin{split}
V_{x}(\mathbf{k}) &= - \frac{1}{L^3} \sum\limits_{k' \leq k_F} \frac{4\pi e^2}{\epsilon |\mathbf{k} - \mathbf{k}'|^2} \\
&= - \int_{k' \leq k_F} \frac{d^3 \mathbf{k}'}{(2\pi)^3} \frac{4\pi e^2}{\epsilon |\mathbf{k} - \mathbf{k}'|^2} \\
&= - \frac{e^2 k_F}{\pi \epsilon} \int_{-1}^1 d\cos\theta \int_0^1 dx \frac{x^2}{x^2 + y^2 - 2xy \cos\theta} \\
&= -\frac{2e^2 k_F}{\pi \epsilon} f_{3D}(\frac{k}{k_F})
\end{split}
\end{equation}
where $x = k'/k_F$, $y = k/k_F$ and
\begin{align}
f_{3D}(x) = \frac{1}{2} + \frac{1-x^2}{4x} \ln \Big| \frac{1+x}{1-x} \Big|.
\end{align}
At finite $T$, the exchange potential is evaluated numerically
\begin{equation}
\begin{split}
V_{x}(\mathbf{k}) &= - \int_{k' \leq k_F} \frac{d^3 \mathbf{k}'}{(2\pi)^3} \frac{4\pi e^2}{\epsilon |\mathbf{k} - \mathbf{k}'|^2} f(\varepsilon_{\mathbf{k}'}) \\
&= -\frac{e^2}{\pi \epsilon} \int_{-1}^1 dz \int_0^{k_F} dk'\frac{(k')^2}{(k')^2 + k^2 - 2k'k z } f(\varepsilon_{\mathbf{k}'}).
\end{split}
\end{equation}
$T_c$ of 3DEG spin polarized state, calculated by finite-temperature SCHF is shown in Fig.~\ref{fig_Tc_3D}(c).

%

\end{document}